\pdfoutput=1

\documentclass[11pt,twoside,a4paper,cmspaper,final,collab]{cms-tdr}
\def\svnVersion{482390P}\def\cmsCernNoTag{CERN-EP-2018-180}\def\cmsCernDate{\today}\def\cmsMessage{Published in the Journal of High Energy Physics as \href{http://dx.doi.org/10.1007/JHEP11(2018)113}{\doi{10.1007/JHEP11(2018)113.}}}

\begin{document}\cmsNoteHeader{SMP-16-010}

\hyphenation{had-ron-i-za-tion}
\hyphenation{cal-or-i-me-ter}
\hyphenation{de-vices}
\RCS$HeadURL: svn+ssh://svn.cern.ch/reps/tdr2/papers/SMP-16-010/trunk/SMP-16-010.tex $
\RCS$Id: SMP-16-010.tex 478683 2018-10-19 19:05:19Z srappocc $
\newlength\cmsFigWidth
\ifthenelse{\boolean{cms@external}}{\setlength\cmsFigWidth{0.85\columnwidth}}{\setlength\cmsFigWidth{0.4\textwidth}}
\ifthenelse{\boolean{cms@external}}{\providecommand{\cmsLeft}{top\xspace}}{\providecommand{\cmsLeft}{left\xspace}}
\ifthenelse{\boolean{cms@external}}{\providecommand{\cmsRight}{bottom\xspace}}{\providecommand{\cmsRight}{right\xspace}}

\newcommand{\NNPDF} {{{NNPDF\,3.0}}\xspace}
\newcommand{\LHAPDF} {{{LHAPDF}}\xspace}

\cmsNoteHeader{SMP-16-010}
\title{Measurements of the differential jet cross section as a function of the jet mass in dijet events from proton-proton collisions at $\sqrt{s} = 13\TeV$}

\date{\today}

\abstract{

Measurements of the differential jet cross section are presented as a function
of the jet mass in dijet events, in bins of jet transverse momentum, with and without
a jet grooming algorithm. The data have been recorded by the CMS Collaboration
in proton-proton collisions at the LHC at a center-of-mass energy of 13\TeV and
correspond to an integrated luminosity of 2.3\fbinv.
The absolute cross sections show slightly different jet
transverse momentum spectra in data and Monte Carlo event generators
for the settings used. Removing this transverse momentum dependence,
the normalized cross section for ungroomed jets is consistent with the
prediction from Monte Carlo event generators for masses below 30\% of
the transverse momentum. The normalized cross section for groomed jets
is measured with higher precision than the ungroomed cross
section. Semi-analytical calculations of the jet mass beyond leading
logarithmic accuracy are compared to data, as well as predictions at
leading order and next-to-leading order, which include parton
showering and hadronization. Overall, in the normalized cross section,
the theoretical predictions agree with the measured cross sections
within the uncertainties for masses from 10 to 30\% of the jet
transverse momentum.
}

\hypersetup{%
pdfauthor={CMS Collaboration},%
pdftitle={Measurements of the differential jet cross section as a function of the jet mass in dijet events from proton-proton collisions at sqrt(s) = 13 TeV},%
pdfsubject={CMS},%
pdfkeywords={CMS, physics, jet cross section}}

\maketitle
\section{Introduction}

The understanding of the evolution of jets in quantum chromodynamics
(QCD) is important to the understanding of physics at hadron colliders.
Jets are involved in nearly every final state in  proton-proton ($\Pp\Pp$)
collisions. In the past, most predictions of jet properties did not
focus on the jet mass $m$, defined as the mass of the sum of the four-momenta of the constituents inside the jet.
However, the jet mass is the simplest observable that is nonzero only
beyond leading order (LO) in perturbation theory for dijet production. Thus, it is more sensitive
to the internal structure and evolution of jets than the jet transverse
momentum $\pt$. The evolution of jets is theoretically described by
parton showering as reviewed in
Refs.~\cite{Hoche:2014rga,Buckley:2011ms}. Experimental measurements
of the jet mass provide
an important test of these theoretical concepts.

Jet mass has been used extensively
to investigate signals of physics beyond the standard model (BSM)
that are reviewed in Refs.~\cite{boost2011,boost2012,boost2013}. In particular,
the jet mass generated by QCD radiation is typically lower than
that generated by heavy hadronically decaying standard model (SM)
particles that have 2- and 3-prong decays, such as top quarks, $\PW$ bosons, $\cPZ$ bosons, or Higgs ($\PH$) bosons.
This observation can be exploited to identify BSM particles that decay into highly Lorentz-boosted SM particles
(boosted objects).

The jet mass generated by QCD arises from the radiation of gluons from
the higher-energy (hard) portion of the jet.
The probability that a gluon is radiated increases as its energy or angle of emission decreases
(soft or collinear radiation, respectively).
Hence, the probability of having $m=0$ vanishes since
many soft and/or collinear gluons will be radiated.
Above $m/\pt\approx 0.1$, there is an overall decrease in the jet mass
spectrum as the dominant contributions come from gluon emissions
with high energy or at large angles. The distribution terminates around
$m/\pt\approx 0.3$ for jets with $R=0.8$ (where $R$ is the distance
parameter of the algorithm). At that point, the
jets become wide enough to start splitting into two separate jets (the splitting
threshold). The appearance of a large peak at low mass is described in Ref.~\cite{mmdt} as a
``Sudakov peak''.
The jet mass is sensitive to soft QCD effects and to
additional $\Pp\Pp$
interactions in the same bunch crossing (pileup)~\cite{SMP-12-019}.
There is further sensitivity in hadron colliders to correlations among
soft gluon emissions from underlying event interactions, and
wide-angle emissions among different jets. While the impacts on global event-wide
observables (such as the event thrust) have been understood for
some time~\cite{Dasgupta:2001sh,Catani:1992ua,Banfi:2001bz,Banfi:2010xy},
the impact on nonglobal jet observables, such as mass, are
difficult to calculate. The technical difficulty arises from the treatment of nonglobal logarithms (NGLs) in
the perturbative next-to-leading logarithmic (NLL) order calculations
of the jet mass that arise from radiation that exits the jet
re-radiating soft gluons back into the jet.
It is currently unknown how to fully account for these NGLs in the perturbative integrals.

The hard part of the jet can be separated from the soft parts using
jet-grooming techniques~\cite{boost2011,boost2012,boost2013}.
These techniques also reduce the effects of pileup,
which is not correlated with the hard part of the jet.
The earliest grooming techniques were unable to completely remove
the soft parts of a jet systematically.
More recently, major improvements have been made to
separate the hard and soft parts of the
jet
without contamination from NGLs~\cite{Dasgupta:2001sh,mmdt,softdrop,Dasgupta:2013via,Dasgupta:2015yua,Larkoski:2015zka}.
The ``mass drop'' (MD) procedure~\cite{Butterworth:2008i} has been
modified into such a theoretically controlled jet-grooming algorithm~\cite{mmdt}.
Its generalization, the ``soft drop'' (SD) algorithm~\cite{softdrop}, has similar properties.
By applying the SD algorithm, soft parts of a jet
are removed, resulting in a dramatically reduced Sudakov peak
in the jet mass distribution.
This also systematically removes the effects that cause NGLs in the
calculation, allowing for direct comparisons of the semi-analytical
NLL resummation calculations (in particular, from the authors of
Refs.~\cite{Frye:2016aiz} and~\cite{Marzani:2017mva}) to Monte Carlo
(MC) simulated predictions and data.
We therefore apply the SD procedure to groom the jets
in this paper. A similar analysis has been
published by the ATLAS Collaboration in Ref.~\cite{atlas_jetmass2}.
Understanding the behavior of the jet mass and the grooming techniques from first principles is
important to further develop boosted object identification algorithms, to gain confidence in the tools
should they uncover BSM physics, and to check that
parton showering algorithms are correct.

In this paper, we present a measurement of the differential jet production cross section
as a function of the jet mass in bins of $\pt$
from dijet events, before
and after the jets are groomed with the SD procedure
(for our choice of parameters the SD and MD procedures are identical).
Comparing the production
cross section with respect to the groomed ($m_{\mathrm{g}}$) and ungroomed
($m_{\mathrm{u}}$) jet mass separately allows us to
gain insight into both the hard and soft physics.

This analysis is similar to that of Ref.~\cite{SMP-12-019}, except we correct both the jet $\pt$ and $m$
for detector effects, and the SD algorithm is used.
We present the double-differential production cross section of dijet events with respect to
the jet \pt and jet mass $m$ (where $m = m_{\mathrm{u}}$ or $m_{\mathrm{g}}$):
\begin{equation}
\frac{\rd^2\sigma}{\rd\pt \, \rd m}  \; (\text{pb}/\GeVns{}^2).
\end{equation}
In addition, we define a normalized distribution:
\begin{equation}
\frac{1}{\rd\sigma/\rd\pt}  \frac{\rd^2 \sigma}{\rd\pt \, \rd m} \; (1/\GeVns{}).
\end{equation}
In both cases, the two jets in the dijet system are included in the distributions.
We have used only the 2.3\fbinv dataset instead of all the available
luminosity, because we present the ungroomed cross section, which is
systematically limited by the pileup uncertainty in higher luminosity
data samples.  

\section{Theoretical predictions}
\label{sec:theory}

There are five theoretical calculations used in this study: three MC
event generator predictions, and two predictions using analytic
resummation merged with fixed order MC simulations.

The three MC event generators used in this study are {\PYTHIA}8, \HERWIGpp, and
\POWHEG. Version 8.212 of {\PYTHIA}8~\cite{Sjostrand:2014zea} is used with underlying event tune
{{CUETP8M1}}~\cite{Khachatryan:2015pea} and \NNPDF parton distribution
functions (PDFs) {{NNPDF30\_lo\_as\_0130}}~\cite{Ball:2014uwa}.
The {{PDF4LHC15\_NLO\_100}} meta-PDF set is used for
uncertainty estimation, along with the {\PYTHIA}8 generator. We use the \LHAPDF~\cite{Buckley:2014ana}
software package for this purpose, following the recommendations of Ref.~\cite{Butterworth:2015oua}.
The \HERWIGpp version 2.7.0 generator~\cite{herwig} is used with tune
{{CUETHS1}}~\cite{Khachatryan:2015pea}.
The \HERWIGpp simulation sample is used for systematic uncertainty estimation and
for generator-level comparisons, as it has a physics model that is different from {\PYTHIA}8 in various aspects like parton showering, hadronisation, multiple-parton interactions, and the set of model parameters tuned to data.
Finally, the \POWHEG 2.0~\cite{Nason:2004rx,Frixione:2007vw,Alioli:2010xd,Alioli:2010xa} generator interfaced with
{\PYTHIA}8 is used
as a next-to-leading order (NLO) perturbative dijet event generator,
supplemented with parton showering and multiparton interactions. The
same underlying event tune
{{CUETP8M1}}~\cite{Khachatryan:2015pea} is used.

Two semi-analytical resummation-based calculations merged with
fixed-order MC simulation are also used. The authors of Ref.~\cite{Frye:2016aiz}
have provided a prediction of the groomed jet mass distribution for dijet events at LO using
\MADGRAPH v2.4.2~\cite{madgraph} and \NNPDF LO PDFs~\cite{Ball:2014uwa}, resummed
at NLL accuracy using soft collinear
effective theory. Nonperturbative effects are included as an
uncertainty in their prediction.
The authors of Ref.~\cite{Marzani:2017mva} have
provided a prediction of the groomed jet mass distribution for dijet events
at NLO using {\textsc{nlojet++}}~\cite{nlojet} and
CT14NLO PDFs~\cite{ct14} with NLL
resummation using perturbative QCD. They include an estimate of
the nonperturbative effects directly, using {\PYTHIA}8.

The semi-analytical predictions are expected to be accurate for
intermediate masses $0.1 < m/\pt < 0.3$, since at low mass nonperturbative effects are
large. Moreover, at high mass the resummed predictions are matched to
fixed-order predictions at LO and NLO, which cannot account
for more than one or two additional gluons, respectively, as would occur within
genuine jets.
For a single gluon emission, there is a kinematic upper limit at $m=\pt R/2$ (where $R=\sqrt{\smash[b]{(\Delta\eta)^2+(\Delta\phi)^2}}$ is the
distance parameter for the jet clustering), from the relativistic kinematics of a $1\to 2$ decay.
However, in the all-orders limit, the approximate kinematic limit is closer to $m=\pt R/\sqrt{2}$.
Consider a particle of energy $E$ and mass $m$ decaying to two massless
particles, each with an energy $E/2$ and separated by an angle $\theta$.
The mass must satisfy $m^2 < E^2\left( 1 - \cos{\theta}\right)/2$. In the small-angle
limit, this would be $m^2 < E^2\theta^2/4$, or $m < E\theta/2$. With more
particles, the stochastic nature of the shower increases this cutoff point.
Thus, a single emission ($1\to 2$) decay will have a lower approximate kinematic
limit than an all-orders ($1 \to$ many) decay.
Reference~\cite{Marzani:2017mva} calculates these thresholds to be
$m/E \approx 0.28$
at LO and $m/E \approx 0.45$ at NLO.

\section{Experimental setup}
\label{sec:expconditions}

The central feature of the CMS detector~\cite{Chatrchyan:2008zzk}
is a superconducting solenoid of 6\unit{m} internal diameter,
providing a magnetic field of 3.8\unit{T}.
Within the superconducting solenoid volume are
a silicon pixel and strip tracker,
a lead tungstate crystal electromagnetic calorimeter (ECAL),
and a brass and scintillator hadron calorimeter (HCAL), each
split in the barrel and two endcap sections.
In the region ${\abs{\eta} < 1.74}$,
the HCAL cells have widths of 0.087 in pseudorapidity ($\eta$)
and 0.087 radians in azimuth ($\phi$).
In the $\eta$-$\phi$ plane, and for ${\abs{\eta} < 1.48}$,
the HCAL cells map onto ${5{\times}5}$ ECAL crystal arrays
to form calorimeter towers projecting radially outwards
from close to the nominal interaction point.
At larger values of $\abs{\eta}$,
the size of the towers increases and
the matching ECAL arrays contain fewer crystals.
Muons are measured in gas-ionization detectors
embedded in the steel flux-return yoke outside the solenoid.
Extensive forward calorimetry complements the coverage provided
by the barrel and endcap detectors.
Electron momenta are estimated by combining the energy measurement in the ECAL with the momentum measurement in the tracker.
In this analysis, the jet rapidity ($y$) is used in the computation of jet
quantities.
A more detailed description of the CMS detector,
together with a definition of the coordinate system used and
the relevant kinematic variables, can be found in Ref.~\cite{Chatrchyan:2008zzk}.

Events of interest are selected using a two-tiered trigger system~\cite{Khachatryan:2016bia}. The first level (L1), composed of custom hardware processors, uses information from the calorimeters and muon detectors to select events at a rate of around 100\unit{kHz} with a latency of less than 4\mus. The second level, known as the high-level trigger (HLT), consists of a farm of processors running a version of the full event reconstruction software optimized for fast processing, and reduces the event rate to around 1\unit{kHz} before data storage.

The event reconstruction is based on
the CMS particle-flow (PF) algorithm~\cite{Sirunyan:2017ulk},
which includes information from all subdetectors, including
charged-particle tracks from the tracking system,
energy deposits in the ECAL and HCAL,
and tracks reconstructed in the muon chambers.
Particles in an event are reconstructed and identified as
electrons, muons, photons, charged hadrons, or neutral hadrons.
Charged hadrons associated with pileup vertices are removed from
consideration. This method is referred to as ``charged-hadron subtraction'' (CHS)~\cite{JME-14-001}.
In our dataset, the average number of pileup interactions per beam crossing is 14.

Jets are clustered using the anti-$\kt$ algorithm~\cite{Cacciari:2008gp}
with the \FASTJET~3.0 software package~\cite{FastJet}, with a distance
parameter of $R=0.8$ (AK8 jets). This distance parameter is chosen to
coincide with the majority of CMS BSM searches using jet substructure techniques.
To reconstruct the groomed jets, the constituents of the AK8 jets are reclustered using the
Cambridge--Aachen algorithm~\cite{CAcambridge,CAaachen} with the
same characteristic radius (CA8) to retain substructure information. The
hierarchical sequence of clustering is then reversed (``declustering''). The
jet grooming algorithm used is the SD algorithm~\cite {softdrop}.
For each declustering step, the ``soft-drop condition'' is applied:
\begin{equation}
\frac{\min(p_{\mathrm{T},i}, p_{\mathrm{T},j})}{(p_{\mathrm{T},i}+p_{\mathrm{T},j})} > z_{\text{cut}} \left(\frac{\Delta R_{ij}}{R_0}\right)^\beta,
\end{equation}
where $p_{\mathrm{T},i}$ and $p_{\mathrm{T},j}$ are the transverse momenta of the $i$th and $j$th
particle at each declustering step, $\Delta R_{ij}$ is the distance between them,
$z_{\mathrm{cut}}$ corresponds to the threshold to separate
the ``soft'' and ``hard'' parts of the jet, and $\beta$ is an angularity
exponent.
If any
declustering step does not satisfy the soft-drop condition, the softer constituent is removed
from consideration by the jet algorithm.
We choose $\beta=0$, $z_{\text{cut}} = 0.1$ and hence $R_0$ is no
longer a relevant parameter.
For this choice of parameters, the SD algorithm is identical to the
variant of the MD algorithm described in Ref.~\cite{mmdt}.
The choice of $\beta=0$ was made for simplicity to agree with most other CMS analyses, 
whereas other values of $\beta$ will be
addressed in future work.  
The measurement is not very sensitive to the value of
$z_{\text{cut}}$, and is chosen to be consistent with
Ref.~\cite{mmdt} and other CMS analyses.

For the reconstructed jets, all PF candidates that survive the CHS procedure are included.
The same reconstruction criteria are also used to create generator-level jets
using generated stable particles (mean path length $c\tau > 1\cm$, no minimum $\pt$ requirement),
removing neutrinos. Simulated pileup
particles are not included in the generator-level jets.
The ungroomed jet mass is taken as the mass of the four-momentum sum of the constituents of the
AK8 jet without grooming applied. The groomed jet mass is taken as the mass of the four-momentum sum
of the constituents of the AK8 jet with the SD algorithm applied. For
this measurement, the $\pt$ of the ungroomed AK8 jet is used for both
the groomed and ungroomed jet cross sections, since it has been
shown that the groomed jet $\pt$ is not collinear-safe for the SD
algorithm chosen~\cite{Marzani:2017mva}.

Since only charged hadrons arising from pileup interactions are removed with
the CHS procedure, corrections based on the jet area~\cite{jetarea_fastjet_pu} are applied to the jet
energies to remove the energy contribution
of neutral hadrons from pileup.
The area-based residual
subtraction is not applied to the jet mass because it is observed~\cite{Khachatryan:2016kdb} that
this overcorrects the mass of hadronically decaying $\PW$ bosons in a
$\ttbar$ control sample.
Further jet energy scale (JES) corrections are applied
as functions of $\eta$ and $\pt$, derived from simulation.
Data-to-simulation correction factors are also applied, using the methodology of Ref.~\cite{Khachatryan:2016kdb}.
These corrections are also applied to the jet mass
by scaling the mass by the same factor used for the $\pt$ in the four-momentum of the jet.

To account for the poorer jet energy resolution (JER) in the data with
respect to 
MC simulation, the reconstructed simulated jets are smeared
compared to generator-level jets by the amounts listed in
Ref.~\cite{Khachatryan:2016kdb}. The typical values of the JER are
around 5--8\%.
After these corrections,
the JES and JER in MC simulation agree with those in data for both groomed
and ungroomed jets.

\begin{figure}[htb!]
\centering
\includegraphics[width=0.495\textwidth]{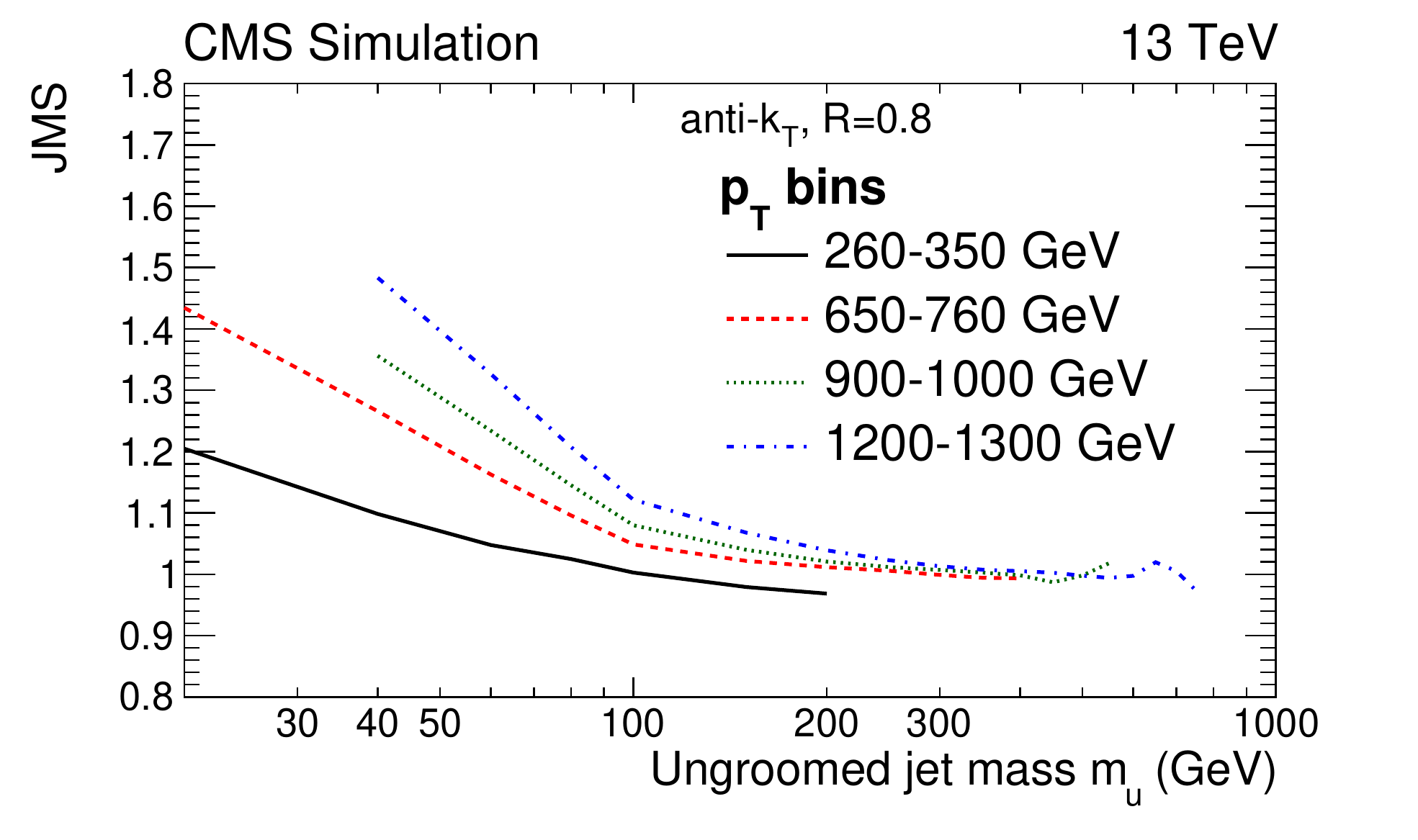}
\includegraphics[width=0.495\textwidth]{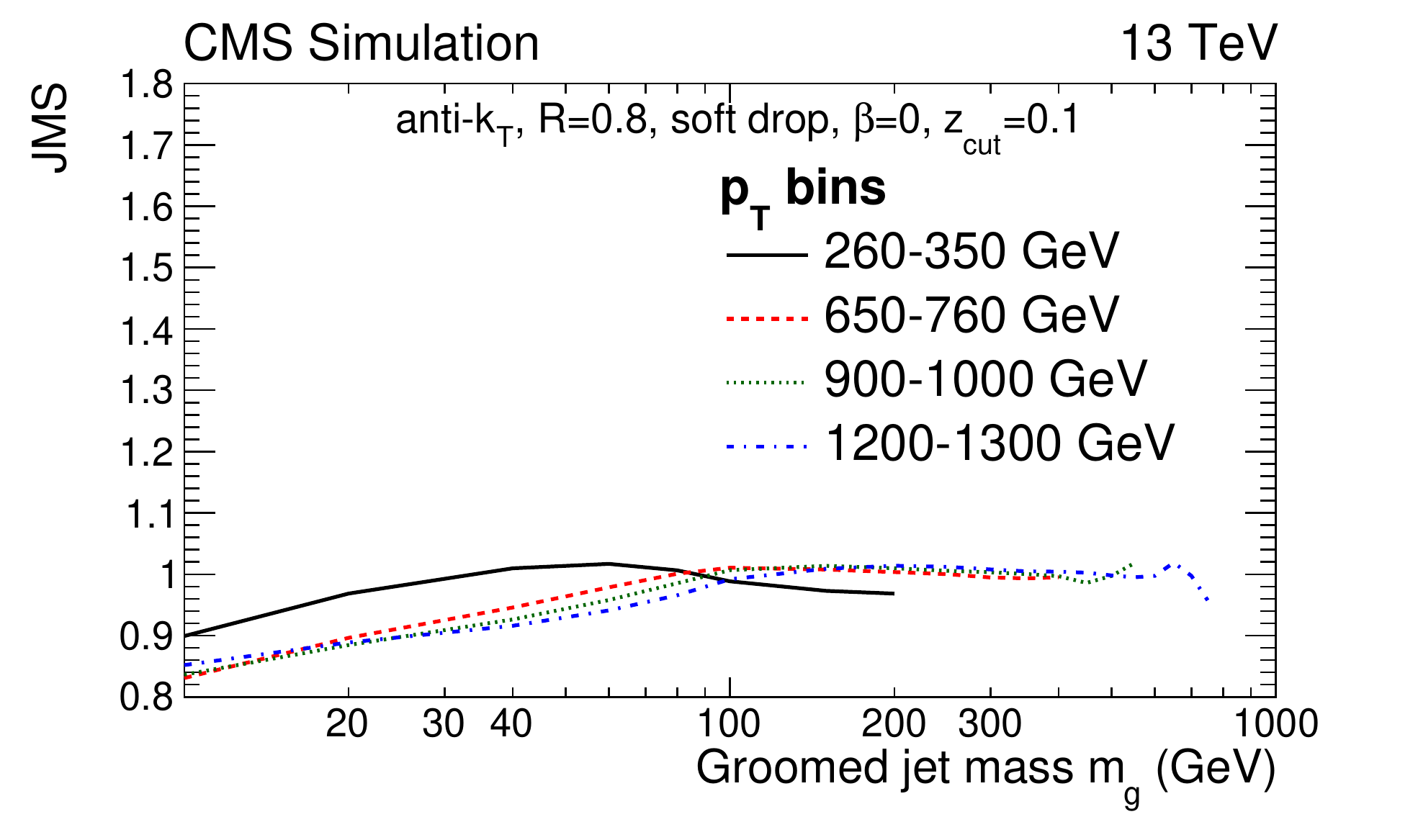}
\caption{JMS in simulation (mean of a fit to $m_{\text{reco}}/m_{\text{gen}}$) for ungroomed (left)
  and groomed (right) jets in different generated ${\pt}$ bins, as a
  function of the generated mass, with our selection. The correction to match the data is
  $0.98 \pm 0.01$. Uncertainties are not shown, for clarity, although
  statistical fluctuations at high masses near the splitting threshold
  for each $\pt$ bin are not negligible.}
\label{fig:jms_allbins}
\end{figure}

\begin{figure}[htb!]
\centering
\includegraphics[width=0.495\textwidth]{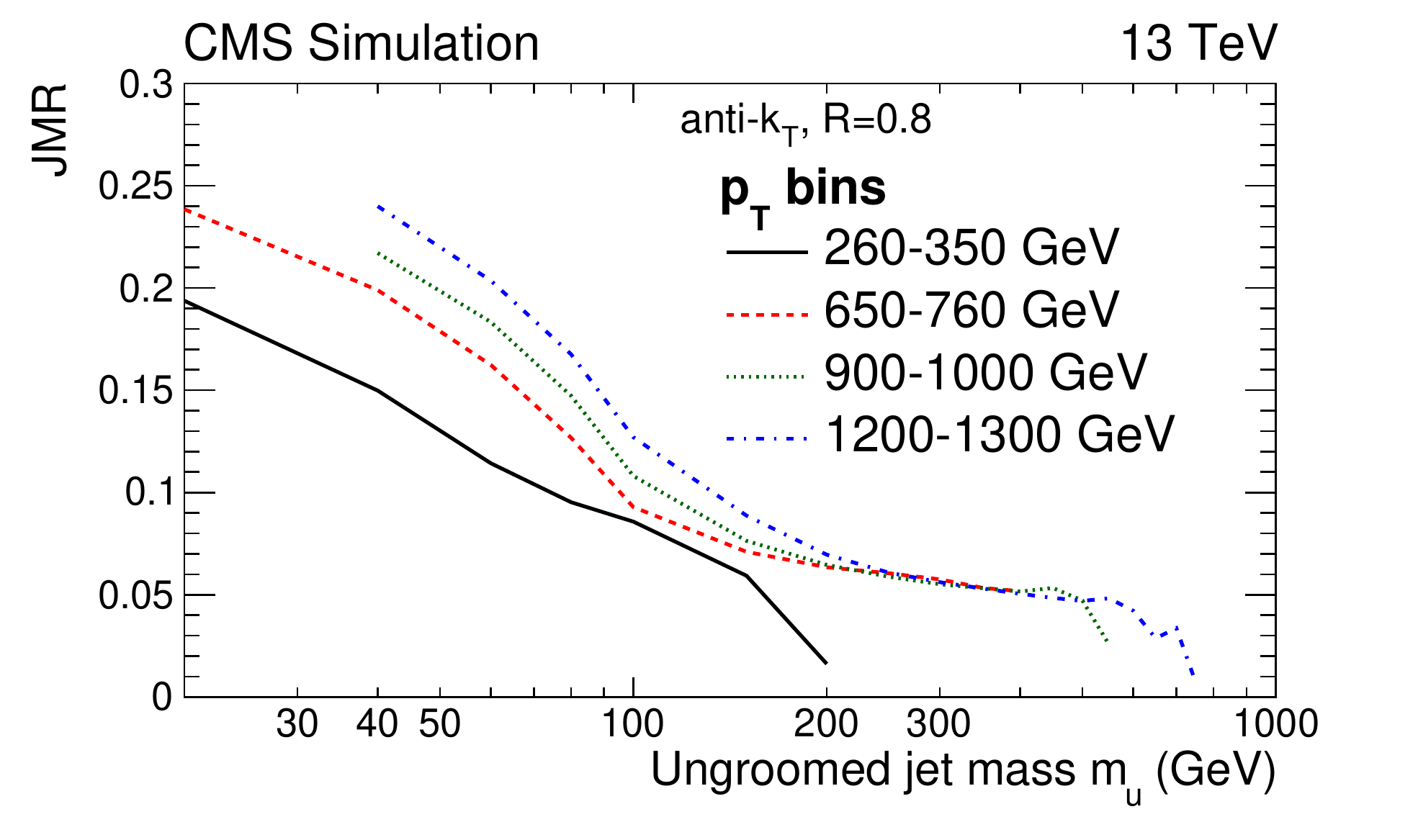}
\includegraphics[width=0.495\textwidth]{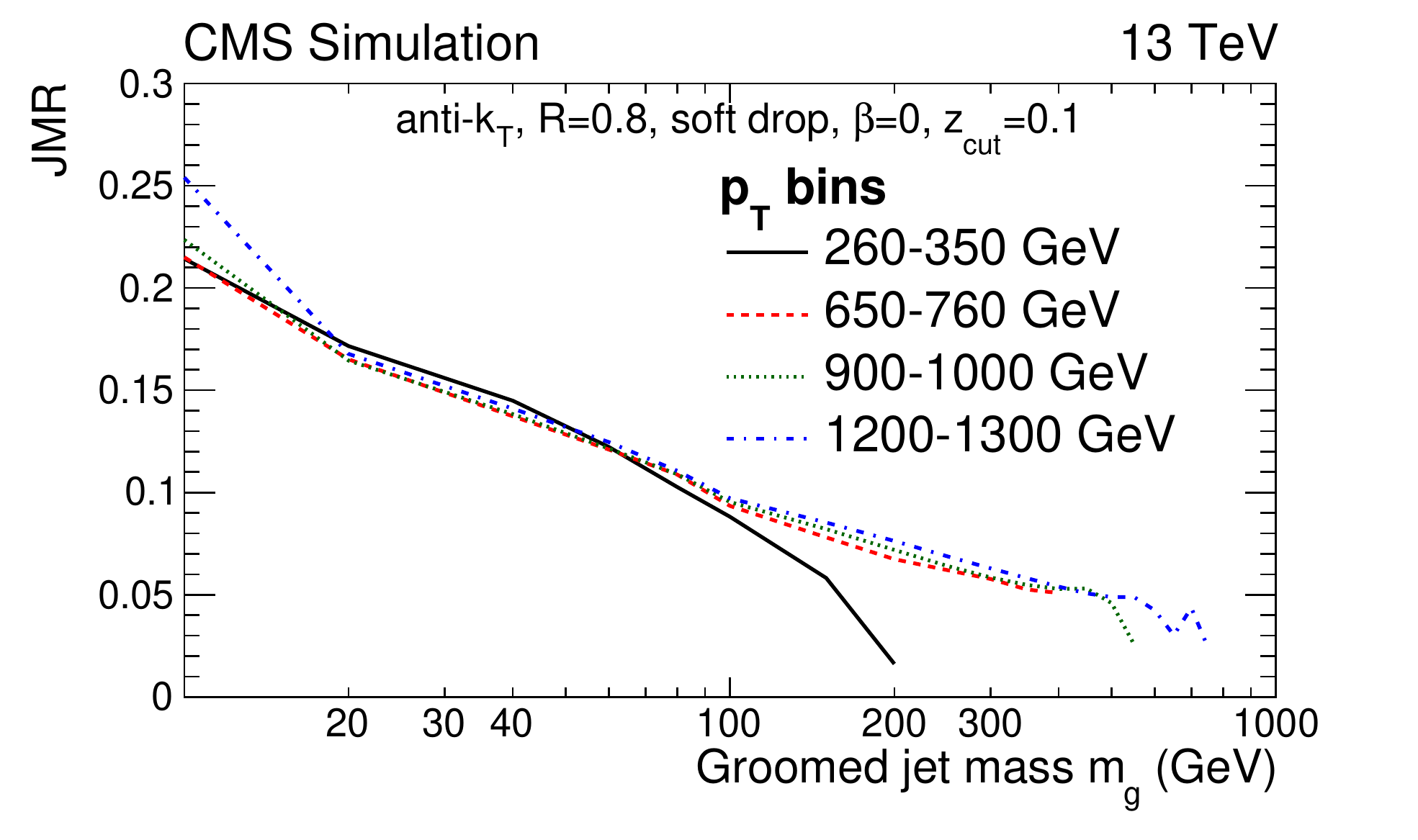}
\caption{JMR in simulation (width of a fit to
  $m_{\text{reco}}/m_{\text{gen}}$) for ungroomed (left) and groomed (right) jets in
  different generated $\pt$ bins, as a function of the generated
  mass, with our selection. The correction to match the data is $1.07
  \pm 0.12$. Uncertainties are not shown, for clarity, although
  statistical fluctuations at high masses near the splitting threshold
  for each $\pt$ bin are not negligible.
} 
\label{fig:jmr_allbins}
\end{figure}

The jet mass can also have differen t scales and resolutions in experimental data
and in MC simulation. 
In the MC simulation, the jet mass scale (JMS) and jet mass resolution (JMR) 
are respectively determined by comparing the means and widths of distributions of
the jet mass at the reconstructed and generator
levels ($m_{\text{reco}}/m_{\text{gen}}$). The JMSs in MC simulation
for groomed and ungroomed jets with our selection
are shown in Fig.~\ref{fig:jms_allbins} as a function of the jet mass for various $\pt$
bins, and similar distributions are shown for the JMR in
Fig.~\ref{fig:jmr_allbins}.
In these figures, finite-size effects of the jet can be observed at
the highest masses in the lowest $\pt$ bin. 
The JMS and JMR are accounted for directly in a detector response parameterization
along with corrections to the MC simulation as to match data. The latter
are derived using the
approach outlined in Ref.~\cite{B2G-16-004}, except we use the SD
algorithm to groom the jets, which is based on a fit of the jet mass from hadronically decaying
$\PW$ bosons in $\ttbar$
events where one $\PW$ boson decays hadronically and the other decays leptonically
(the ``lepton plus jets'' channel). The ratio of the fitted means in data
and in MC simulation is the correction to the JMS,
and the ratio of the fitted widths is the correction to the JMR.
This estimate assumes that detector-related uncertainties will be the
same for jets from $\PW$ boson decays and jets initiated by quarks and gluons.
The remaining differences are effects from the physics model, and are thus accounted
for with our physics model uncertainty.
The JMR and JMS corrections
in the 2015 run of the LHC  (``Run 2'') are measured to be $1.07 \pm
0.12$ and $0.98 \pm 0.01$, respectively, using the methods of Ref.~\cite{B2G-16-004}. 
In the 2010-2012 run of the LHC  (``Run 1''), they were observed to be $1.11\pm 0.09$ and $1.017 \pm 0.006$,
respectively~\cite{JME-13-006}.
In addition to smearing the entire four-vector (including
the $\pt$ and the mass) of the jet by the JER, we also correct the jet mass
distribution in the MC simulation to match the data in the $\PW$
reference sample.

The data used in this analysis are from $\Pp\Pp$ collisions collected in 2015 with 50\unit{ns}
bunch spacing, as in Ref.~\cite{SMP-15-007}, as well as with 25\unit{ns}
bunch spacing.
The integrated luminosities of the 25 and 50\unit{ns} samples correspond
to approximately 2.3 and 0.07\fbinv, respectively.
The data were collected with single-jet triggers of varying
$\pt$ thresholds
as in the inclusive jet analysis from
Ref.~\cite{SMP-15-007}. As in that paper, the ensemble of events is
constructed such that any given trigger only contributes to a specific
$\pt$ region, where it is fully efficient.
However, since we are using a different $R$-parameter for the jet
clustering algorithm
(we use $R=0.8$ as opposed to $R=0.4$ and $0.7$),
our thresholds, shown in Table~\ref{thresholds}, are slightly different than those from the
inclusive jet analysis ~\cite{SMP-15-007}.
The same $\pt$ thresholds are used for both 50 and 25\unit{ns} data.
In the collisions with 50\unit{ns} bunch spacing, the trigger prescales were lower,
and thus this dataset increases the number of events collected for
$\pt < 460$\GeV by 10\%
compared to the 25\unit{ns} run alone,
where the prescales are considerably higher. There are 43.9 million events in
the 25\unit{ns} sample and 4.4 million events in the 50\unit{ns} sample.

\begin{table}[htbp]
\begin{center}
\topcaption{List of trigger thresholds used for each high-level trigger path. The
ungroomed jet $\pt$
ranges are shown in the second column. Events are selected such that
exactly one fully efficient trigger is used for each $\pt$ bin.\label{tab:trig}}
\begin{tabular}{cc}
Trigger $\pt$ threshold ({\GeVns}) & $\pt$ range ({\GeVns}) \\
\hline
  80	& 200--260 \\
  140	& 260--350 \\
  200	& 350--460 \\
  260	& 460--550 \\
  320	& 550--650 \\
  400	& 650--760 \\
  450	& $>$760  \\
\hline
\end{tabular}

\label{thresholds}
\end{center}
\end{table}

\section{Analysis strategy}

The measurement of the dijet cross section is corrected for detector
effects using an unfolding technique. The detector response
is parameterized as a four-dimensional response matrix representing
the migrations from reconstructed $m$ and $\pt$ to generated $m$ and $\pt$.  
This response matrix is used to correct the measured
distribution, and the resulting unfolded distribution is compared to
various theoretical models.

The events are required to have at least two jets, without an explicit third jet veto,
which allows for an easier comparison to semi-analytical calculations. To ensure a
high-purity dijet-like sample, we require that the asymmetry between the leading and
subleading jet transverse momenta, defined as ${\pt}_1$ and ${\pt}_2$ respectively, satisfies
$({\pt}_1 - {\pt}_2) /  ({\pt}_1 + {\pt}_2) < 0.3$ and the difference in azimuthal angles
satisfies $\Delta (\phi_{1} - \phi_{2}) > \pi/2$.
In addition to ensuring a dijet topology, these criteria reduce the
number of jets from detector noise. Both jets in the dijet system are included in the measurement.

All reconstructed AK8 jets are required to have $\pt > 200$\GeV and
rapidity $\abs{y} < 2.4$, and jet identification
requirements that remove detector noise outlined in Ref.~\cite{CMS-PAS-JME-16-003}.
After the identification and topological requirements, the number of jets from detector noise in this sample is negligible. 
Generator-level AK8 jets are required to have $\abs{y} < 2.4$ and a very
low $\pt$ selection ($>$50\GeV) to ensure an accurate estimate of the migration
of events between $\pt$ bins, but the measurement is presented only
for $\pt > 200$\GeV at the generator level. The $\pt$ asymmetry and $\Delta \phi$ selections are
also applied at the generator level, but the $\pt$ asymmetry is a
slowly varying function of $\pt$, so this does not strongly affect the
$\pt$ migration.

The bins in ungroomed jet $\pt$ are chosen so that each bin is
populated by exactly one trigger, with binning as shown in Table~\ref{tab:trig} (i.e., the trigger with a threshold of 80\GeV will be used to populate the 200-260\GeV bin, etc).
Above the threshold of the lowest unprescaled trigger at
$\pt = 760$\GeV,
more uniform $\pt$ binning is used.
The lower bin boundaries are 200, 260, 350, 460, 550, 650, 760, 900, 1000, 1100, 1200, $>$1300\GeV.
The choice of jet mass
bins is a trade-off among the approximate detector
resolution of $\sim$10\GeV~\cite{Khachatryan:2016kdb}, the purity (the fraction of events in a
reconstructed bin that are generated in the same bin),
and the stability (the fraction of events in a generated bin that are
reconstructed in the same bin).
The lower bin boundaries are 10, 20, 40, 60, 80, 100\GeV and then increase by 50\GeV
beyond 100\GeV.
The same binning is used for both the reconstructed and generated
jet mass.
There are, however,
kinematic regions that have insufficient accuracy for the jet mass measurement.
This inaccuracy can arise from the finite spatial resolution of the detector or from excessive
pileup contamination.
These regions in $\pt$ and $m$ are removed by requiring that the uncertainty in
a Gaussian fit to
the distribution of $m_{\text{reco}}/m_{\text{gen}}$ in MC simulation is
less than 50\% for that region.
We do not consider the measurements below those thresholds.
For ungroomed jets, the dominant effect is pileup contamination,
and we do not consider
jet masses below 20\GeV when $\pt < 760$\GeV, or below 40\GeV when
$\pt > 760$\GeV. For groomed jets, the dominant effect is the
finite spatial resolution of the detector, and we do not consider
jet masses below 10\GeV for all $\pt$ bins.

The unfolding procedure of D'Agostini with early stopping is applied~\cite{D'Agostini:1994zf,Richardson:72,Lucy:1974yx}.
The detector response is from MC simulation, and parameterized in bins of
ungroomed jet $\pt$ and ungroomed jet mass for the ungroomed case, and in bins
of ungroomed jet $\pt$ and groomed jet mass for the groomed case.
The central value of the response matrix
is constructed using {\PYTHIA}8. The uncertainty in the response matrix is estimated by
varying the systematic uncertainties described below, rederiving the matrix for each case, and applying
the modified response matrix
to the reconstructed data, taking the difference in the unfolded distribution as the resultant uncertainty.

The number of iterations in the early stopping procedure has been optimized as follows. The response matrix is used to
unfold the {\PYTHIA}8 MC simulation varying the number of iterations in the regularization.
A simple matrix multiplication is then used to refold the unfolded MC simulation.
The resulting refolded distribution is compared to the original
reconstructed distribution and a $\chi^2$ probability is calculated.
The optimal number of iterations is determined as the lowest number such that the $\chi^2$ probability exceeded 99\%.
We found $n=3$ to be optimal.
Varying the number of iterations did not
lead to a significant change in the results.  

The uncertainties in the unfolded result due to statistical
uncertainties in the MC simulation are nontrivial to estimate
in the unfolding procedure because the events have different weights,
so the statistical uncertainty is not correctly estimated by simply
adding the uncertainties from
weighted events in quadrature.
Instead of standard methods, we employ a ``jackknife resampling'' technique (a linear
bootstrapping method)~\cite{10.2307/2237363,10.2307/2983696}, where we
create 10 samples, each with 90\% of the MC events (such that each
event is excluded from exactly one resampling), and unfold
the spectrum. The root-mean-square of this ensemble of distributions, multiplied by
$\sqrt{10/9}$ to account for using 90\% of the events per iteration,
is taken as the statistical uncertainty in the MC simulation.

\label{sec:uncertainties}

There are both theoretical and experimental systematic uncertainties that affect
this measurement, along with uncertainties from the unfolding
procedure. The uncertainties are estimated
by creating response matrices with the systematic effect applied,
and unfolding the reconstructed data to account for systematic migrations.
Unless otherwise specified, the difference between the
systematically varied output and the nominal output is taken to be the
systematic uncertainty for that source.
The uncertainties from different sources are then added together in quadrature.

The experimental uncertainties that affect the
measurement are the JES, JER, JMS, JMR, pileup, and luminosity.
The theoretical uncertainties include uncertainties in the
physics model and PDFs.

The JES uncertainty is evaluated by
varying the jet energy and mass up and down by the appropriate
uncertainties for all of the JES uncertainty sources as described in Ref.~\cite{Khachatryan:2016kdb}.
The JER uncertainty is evaluated by
smearing the jet energy and mass by an additional factor as described in
Ref.~\cite{Khachatryan:2016kdb}. The central value is determined by correcting
the MC simulation to match the JER from data, and the uncertainties are evaluated by
correcting the MC simulation by the JER systematic variations. This is also propagated to
the jet mass because these effects are correlated for jets reconstructed from
PF candidates. The uncertainty due to the finite resolution of the jet angle
is found to be negligible compared to other uncertainties.

As mentioned in Section~\ref{sec:expconditions},
the JMR in simulation is corrected by $1.07 \pm 0.12$ to match
the measured difference between data and MC simulation for the mass of
$\PW$ boson jets in lepton+jets $\ttbar$ events~\cite{Chatrchyan:2011ds}.
The JMS in MC simulation approaches unity for larger
jet $\pt$ and masses. Deviations from unity are included directly
in the response matrix, as described in Section~\ref{sec:expconditions}.

The pileup uncertainty is estimated by reweighting
the pileup distribution of the MC simulation up and down, assuming a change of the minimum bias cross section by 2.7\%~\cite{lumi}.
For the absolute cross section, the luminosity uncertainty of 2.7\% is
also taken as an additional systematic uncertainty.

The PDF uncertainty is
taken from the {\textsc{pdf4lhc15\_nlo\_100}\xspace meta-PDF set using the {\PYTHIA}8 generator.
The physics model uncertainty is evaluated
by comparing a response matrix derived from {\PYTHIA}8 to a
response matrix derived from \HERWIGpp. The difference
between the two results for the two cases is taken as the
systematic uncertainty.

The jet mass distributions in data before unfolding are
compared to {\PYTHIA}8 and \HERWIGpp MC simulations in
Fig.~\ref{fig:PreUnfold1}. 
The uncertainties before the unfolding are evaluated by varying the MC simulation
up and down by each systematic effect described in the previous
paragraphs, and using the sum of the differences to the nominal case
in quadrature as the uncertainty band.
The MC simulation distributions include the theoretical systematic uncertainties,
the experimental systematic uncertainties,
and the statistical uncertainties of the MC samples.
The data distributions include only the statistical uncertainties.
These data distributions are unfolded to obtain the cross section
measurements.

\begin{figure}[htb!]
\centering
\includegraphics[width=0.49\textwidth]{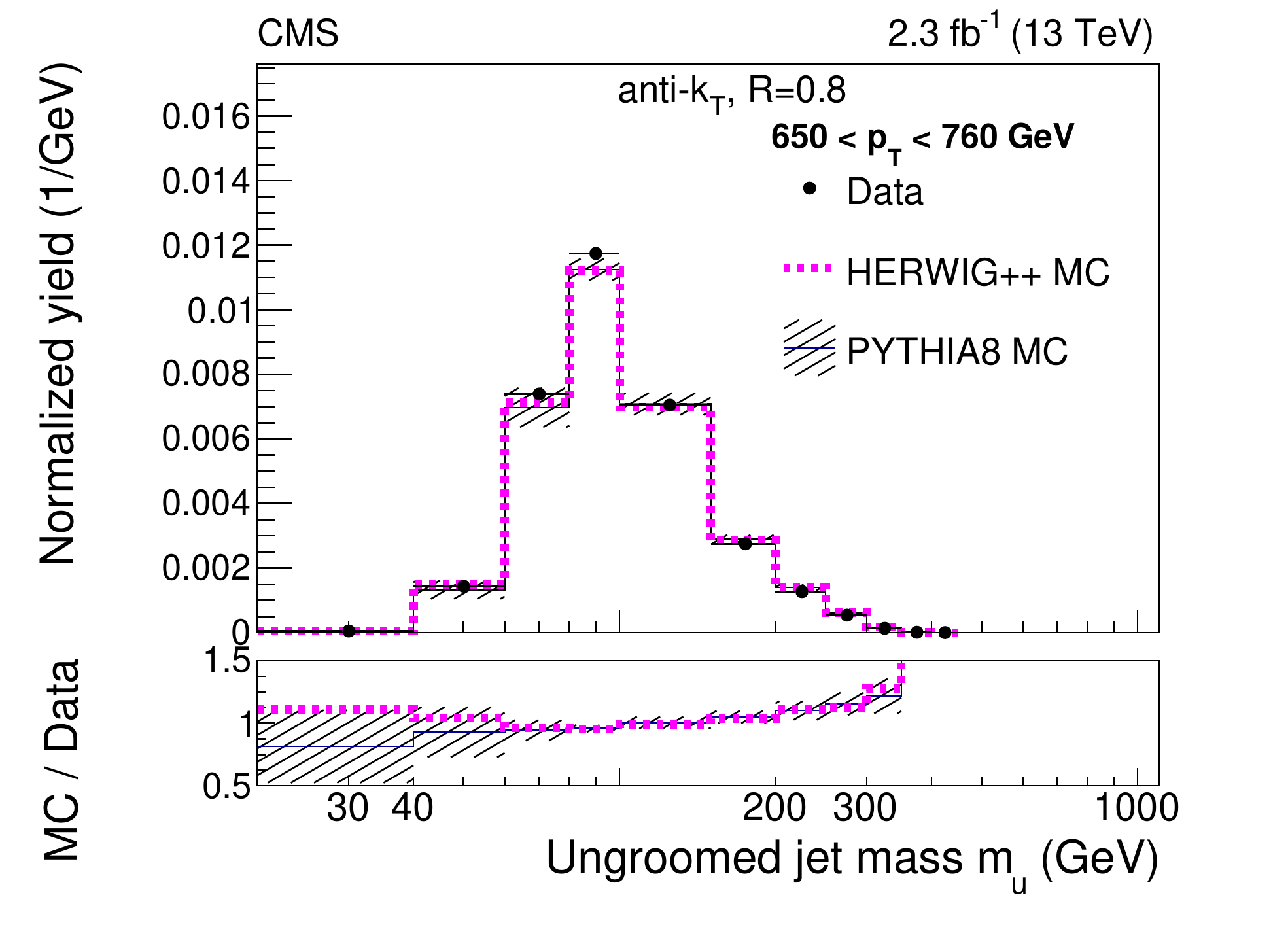}
\includegraphics[width=0.49\textwidth]{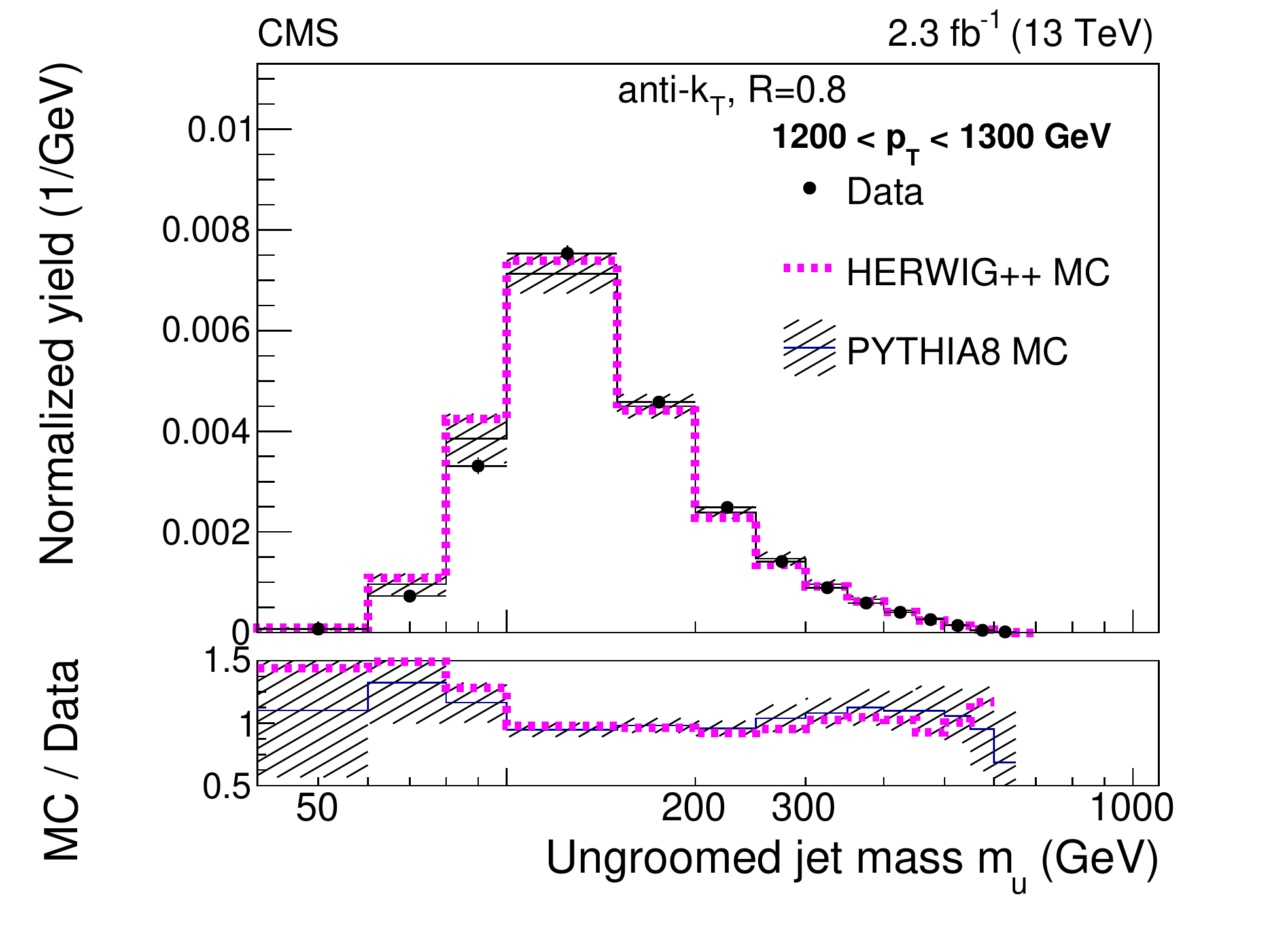}\\
\includegraphics[width=0.49\textwidth]{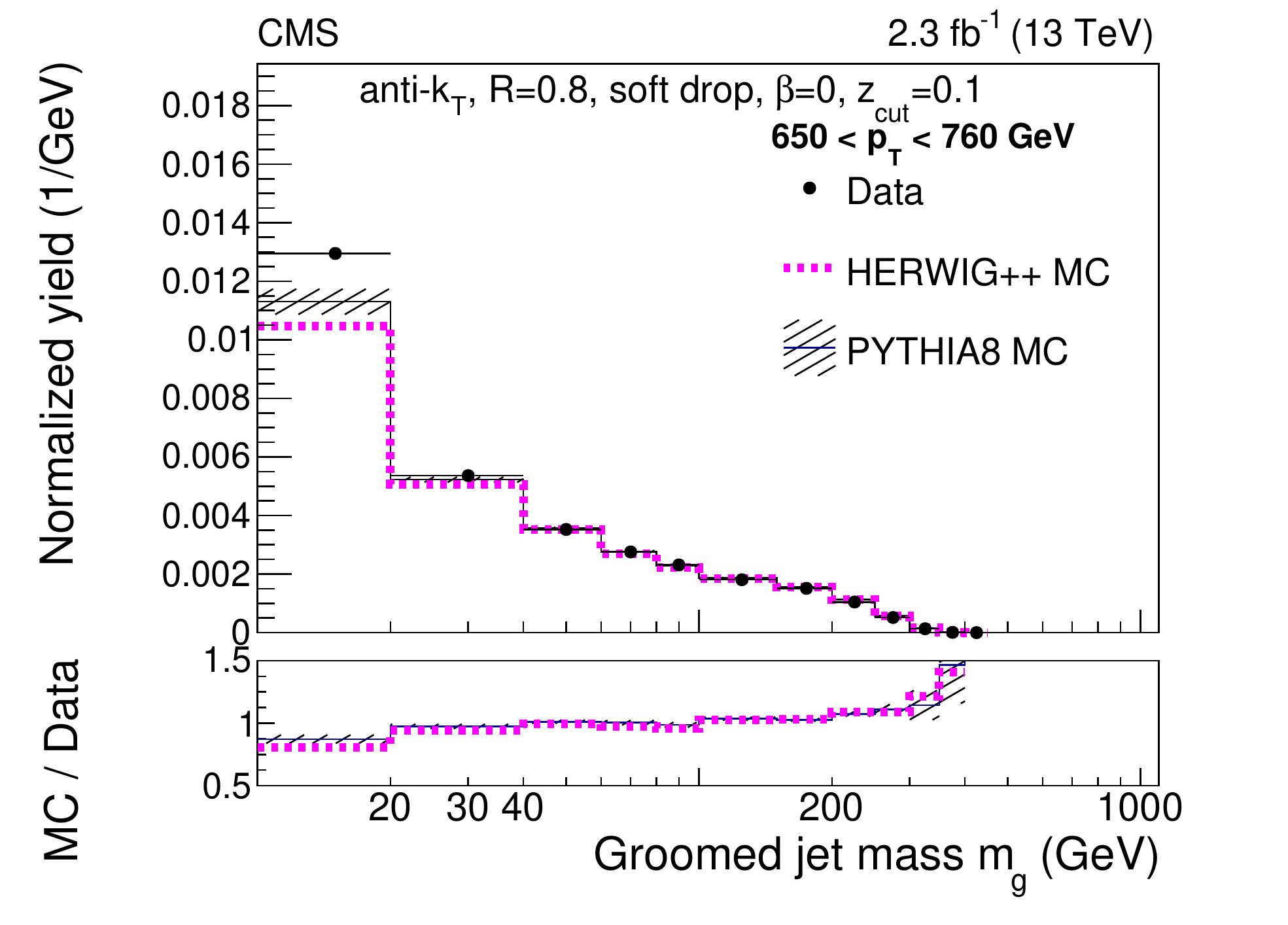}
\includegraphics[width=0.49\textwidth]{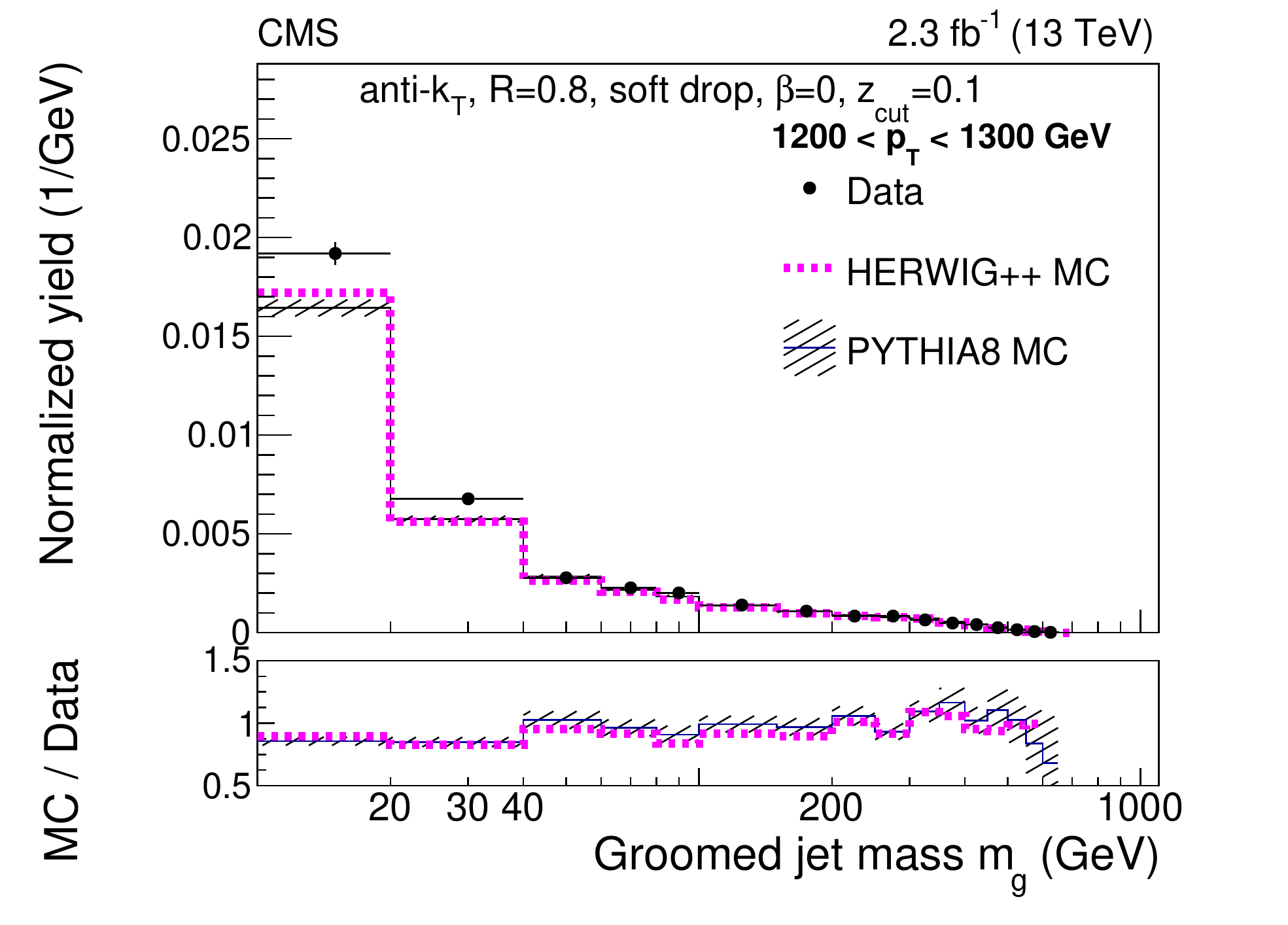}
\caption{Comparison of data to MC simulation for ungroomed jets (top)
  and groomed jets (bottom) for two $\pt$ bins at the detector level. 
The data and their statistical uncertainties are shown by the black points with the error bars.
The {\PYTHIA}8 results
before the unfolding
are shown in the shaded histograms, including experimental and theoretical systematic uncertainties.
The \HERWIGpp results before unfolding are shown without uncertainties as a dashed magenta histogram. }
\label{fig:PreUnfold1}
\end{figure}

Figure~\ref{fig:AbsUnc1} shows the uncertainties in
the absolute cross section for all of the
various sources listed above, while 
Fig.~\ref{fig:Unc1} shows the same uncertainties in
the normalized cross section. The jet energy uncertainties in the normalized cross
sections are smaller than in the absolute cross sections because the distributions
are normalized per $\pt$ bin.
The uncertainties at low jet mass are dominated by
the physics model, JMS, JMR, and pileup uncertainties for the ungroomed jets.
However, the grooming procedure dramatically reduces the physics model uncertainties
by removing the soft components that are difficult to model, as well as the
pileup uncertainty, since pileup mimics soft radiation.
At higher jet mass, the jets are at the threshold of splitting into two, so
the uncertainties increase because of
$\pt$ bin migration. In addition,
the phase space is severely restricted in this regime, so the statistical uncertainty
also grows.

\begin{figure}[htb!]
\centering
\includegraphics[width=0.49\textwidth]{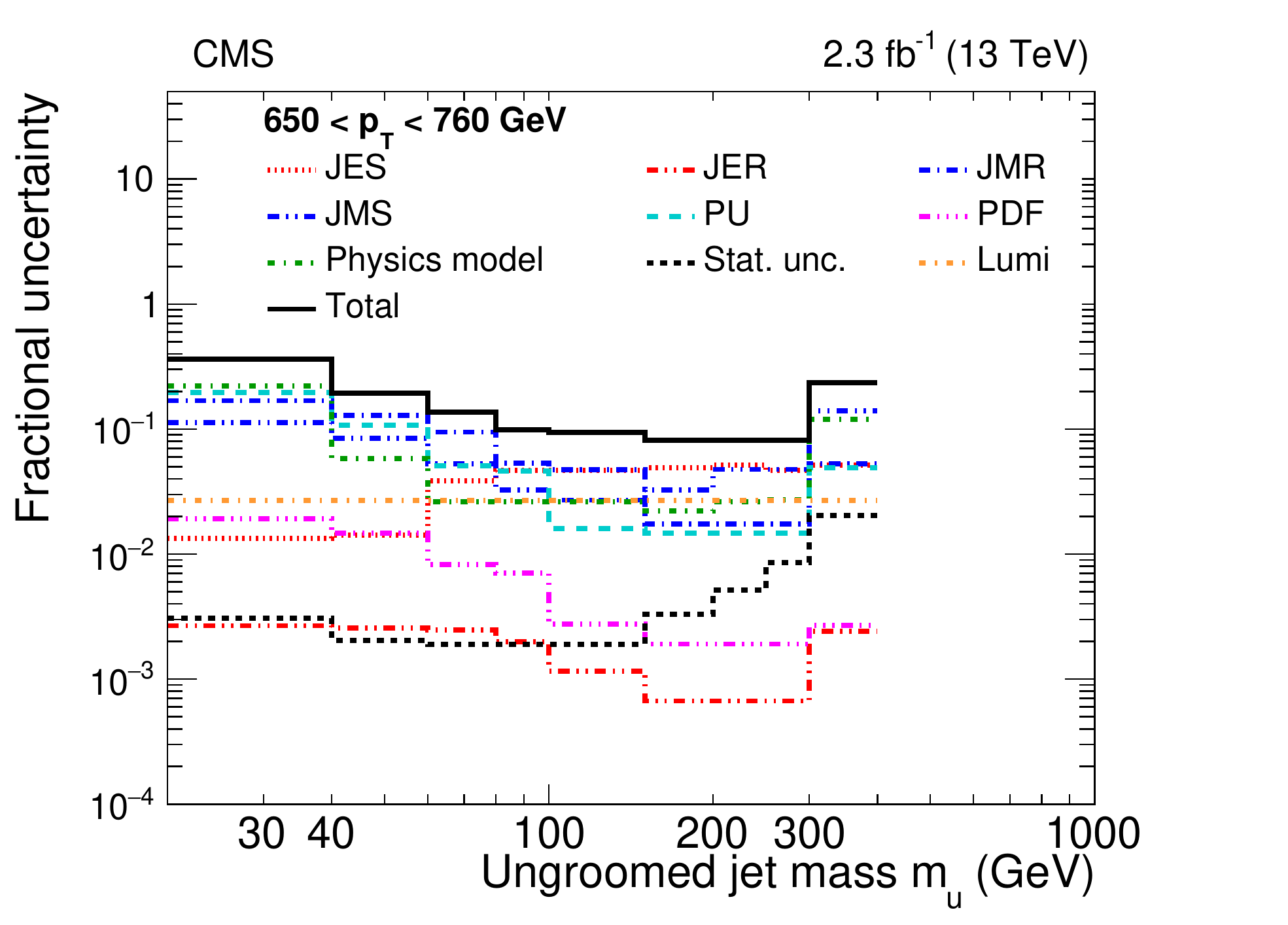}
\includegraphics[width=0.49\textwidth]{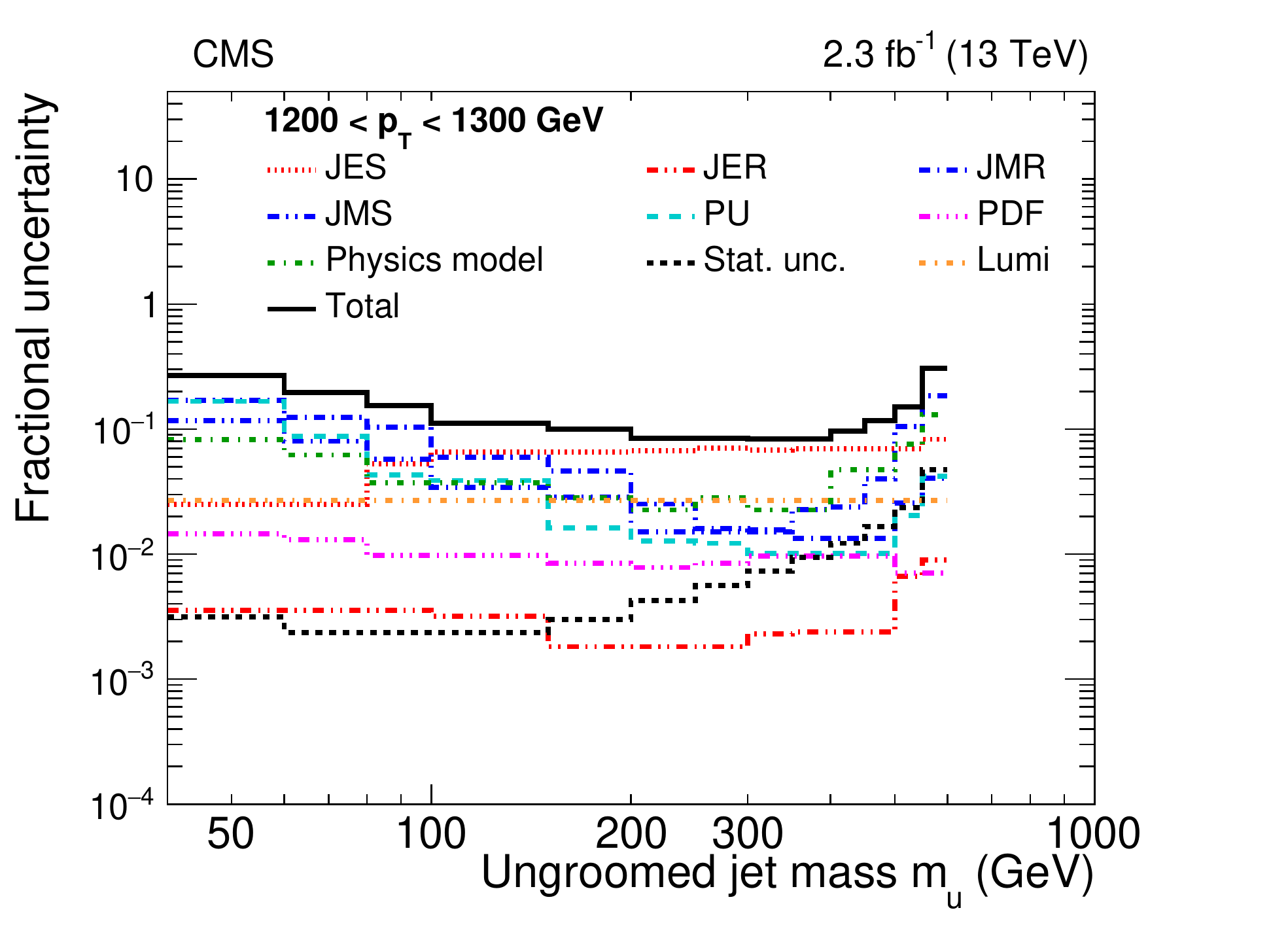}\\
\includegraphics[width=0.49\textwidth]{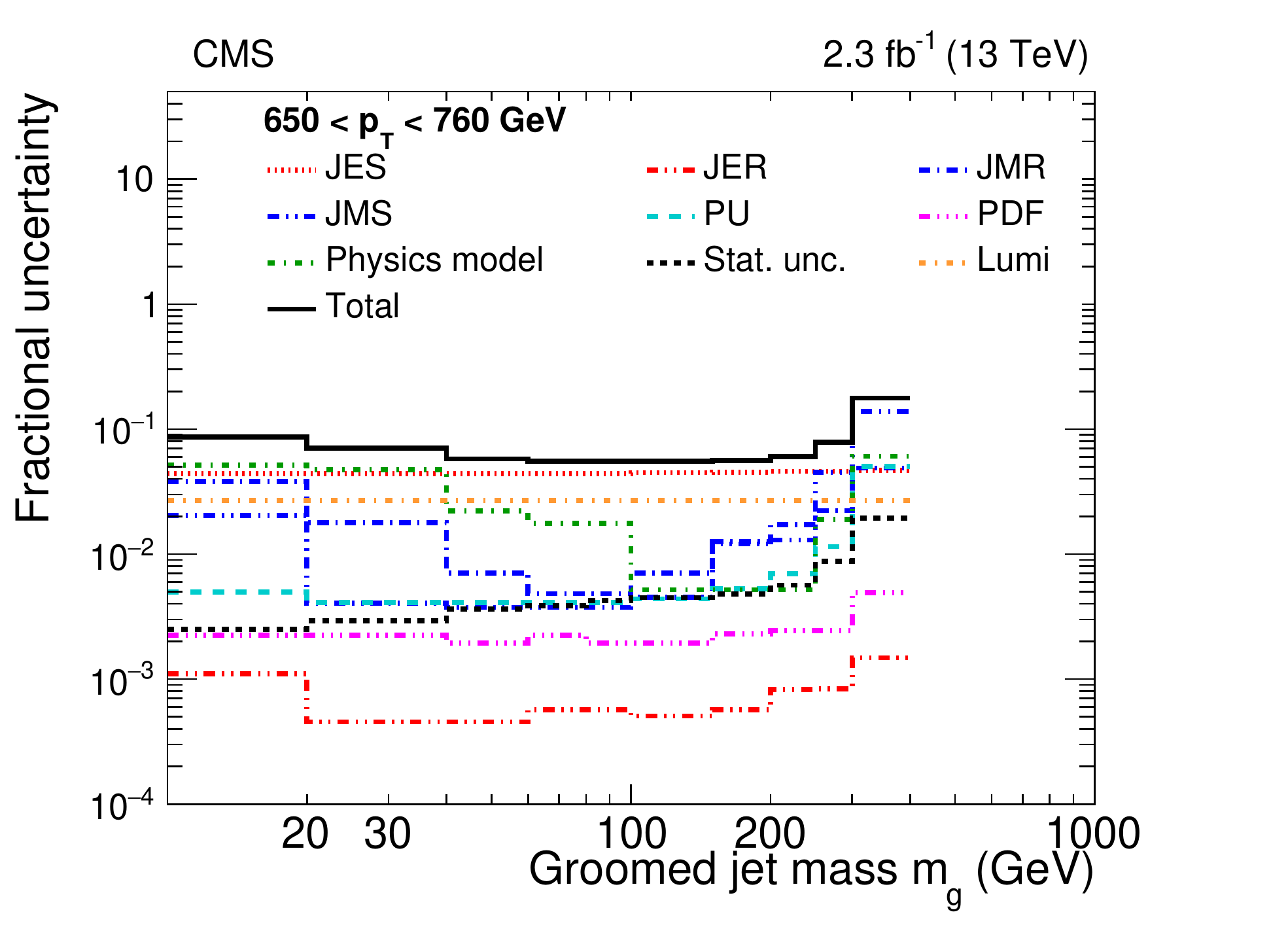}
\includegraphics[width=0.49\textwidth]{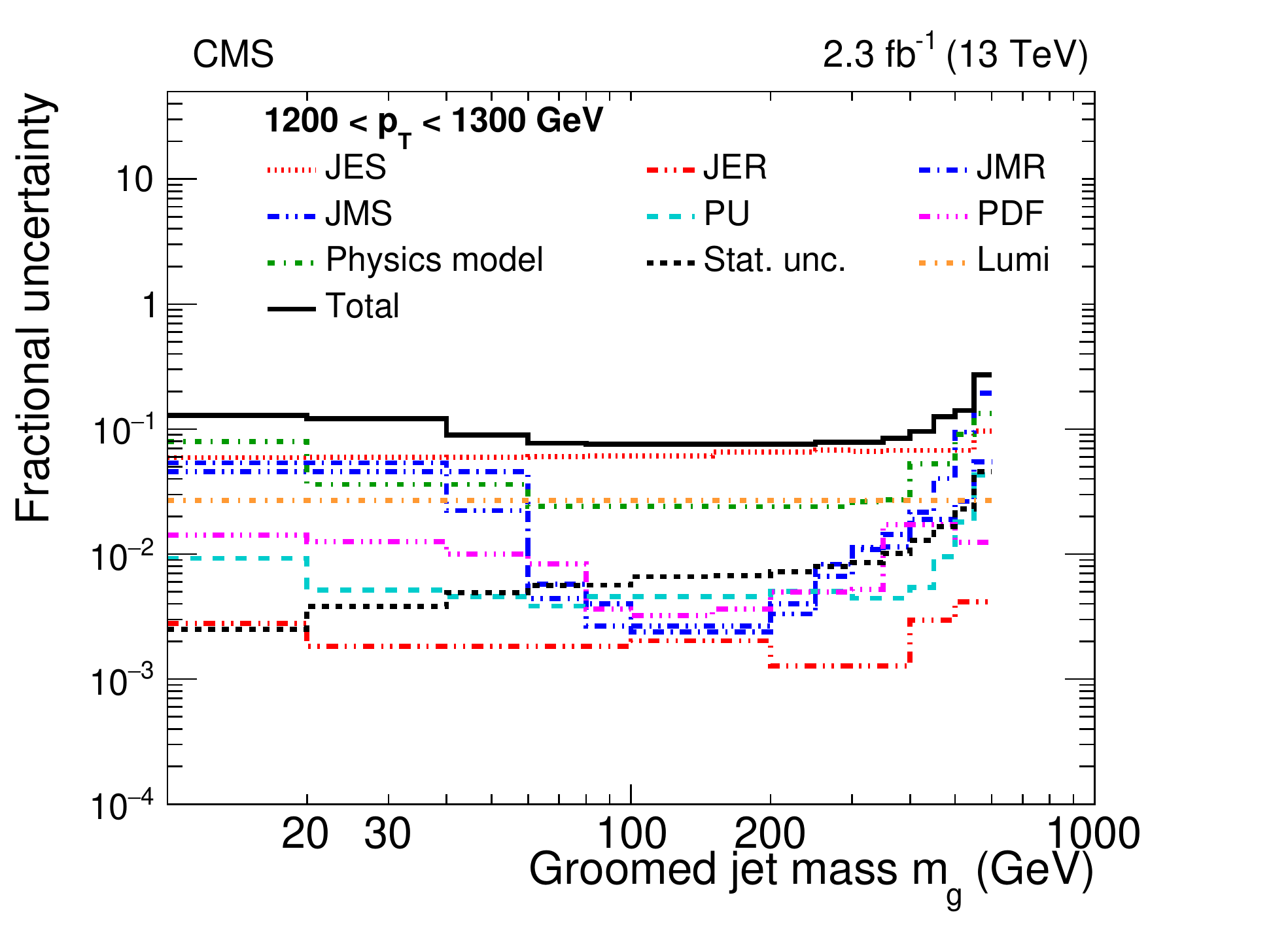}
\caption{Systematic uncertainties in the absolute cross section after unfolding for two $\pt$ bins
  before grooming (top) and after grooming (bottom). The red dotted lines show the JES uncertainties. The red dash-dot-dotted lines show the JER uncertainties. The blue dash-dotted lines show the JMR uncertainties. The blue dash-dot-dotted lines show the JMS uncertainties. The cyan dashed lines show the PU uncertainties. The orange dash-dotted lines show the luminosity uncertainties. The magenta dash-dot-dot-dotted lines show the PDF uncertainties. The green dash-dotted lines show the physics model uncertainties. The black short-dashed lines show the statistical uncertainties.} 
\label{fig:AbsUnc1}
\end{figure}

\clearpage

\begin{figure}[htb]
\centering
\includegraphics[width=0.49\textwidth]{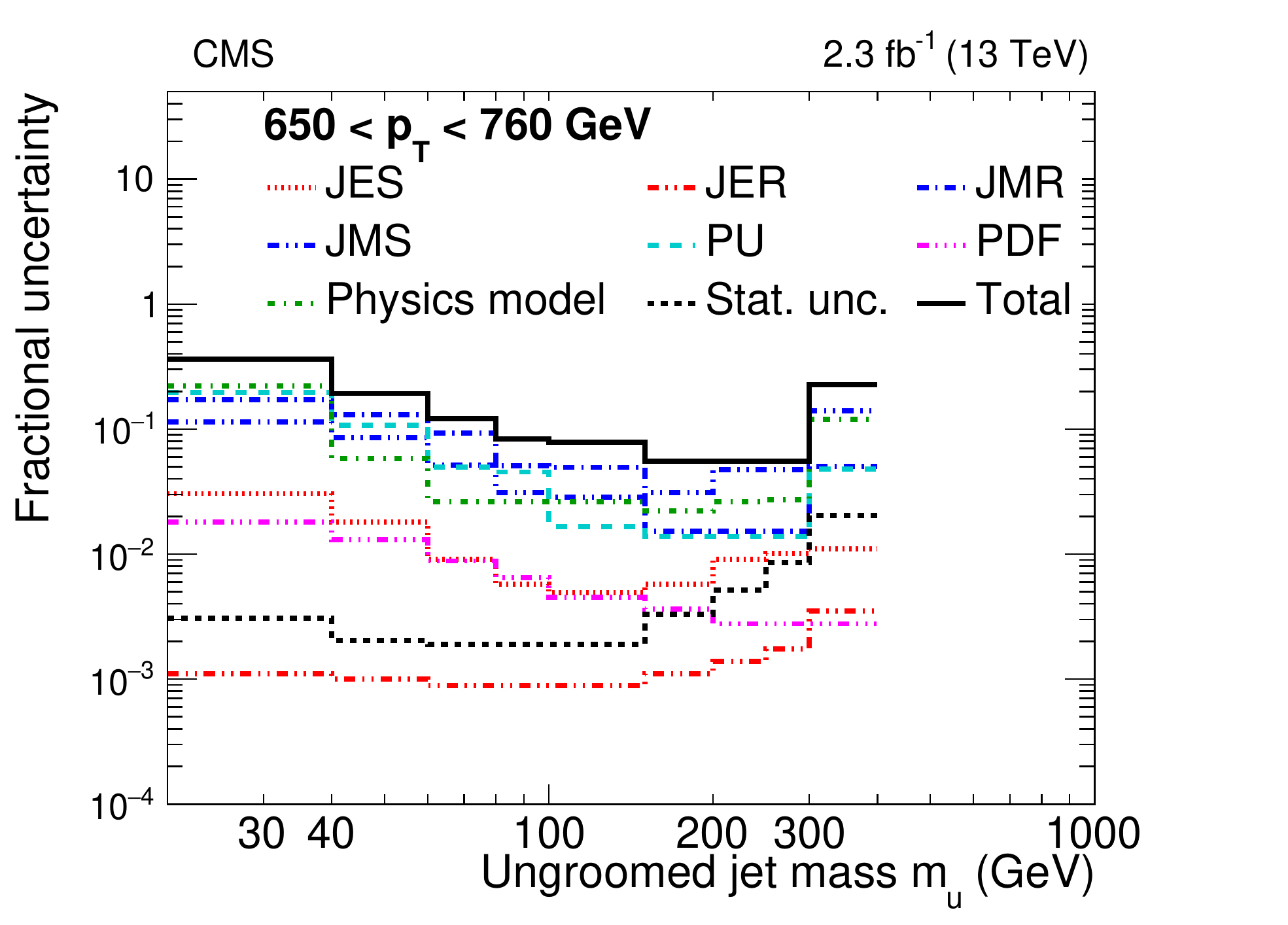}
\includegraphics[width=0.49\textwidth]{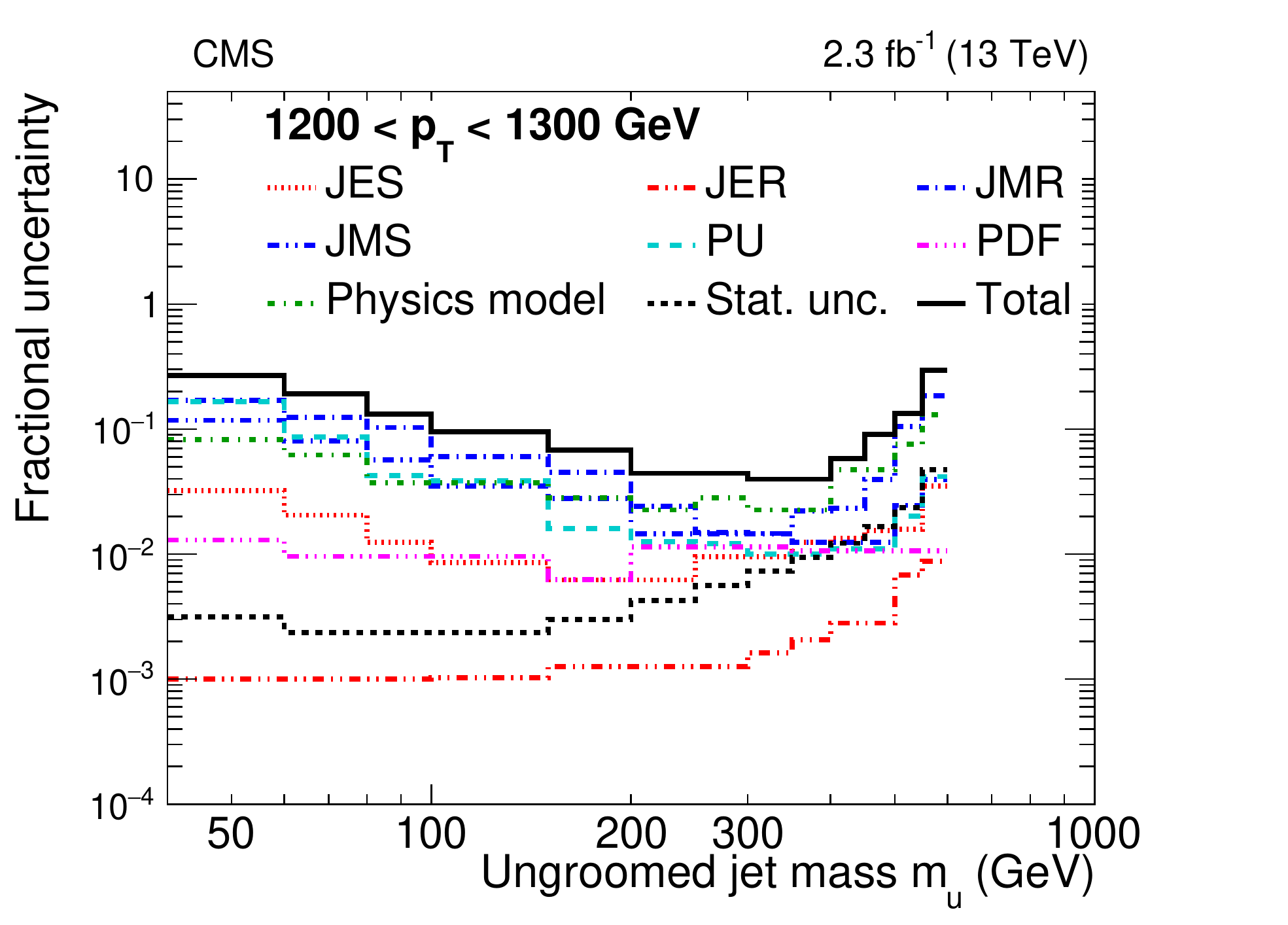}\\
\includegraphics[width=0.49\textwidth]{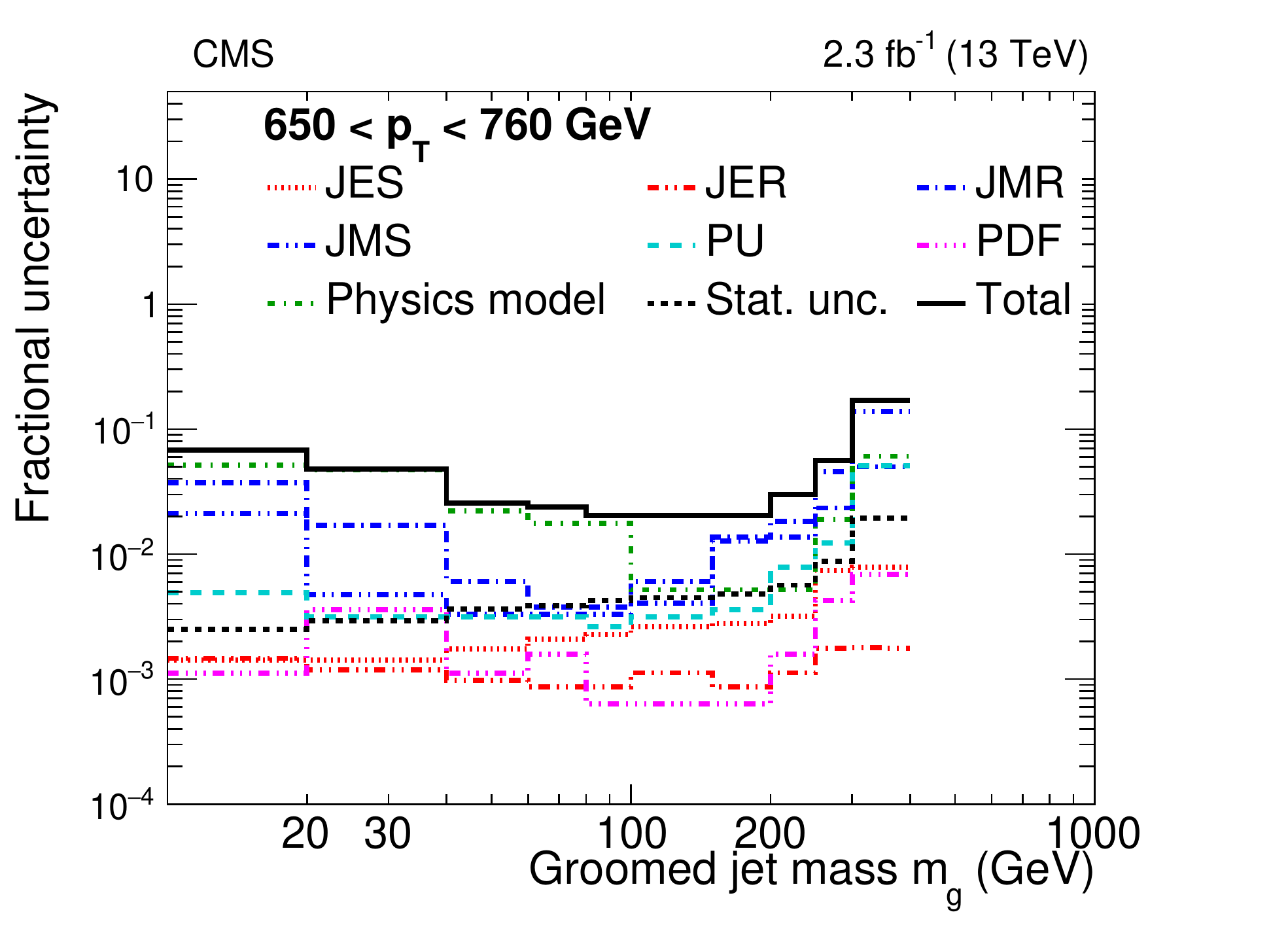}
\includegraphics[width=0.49\textwidth]{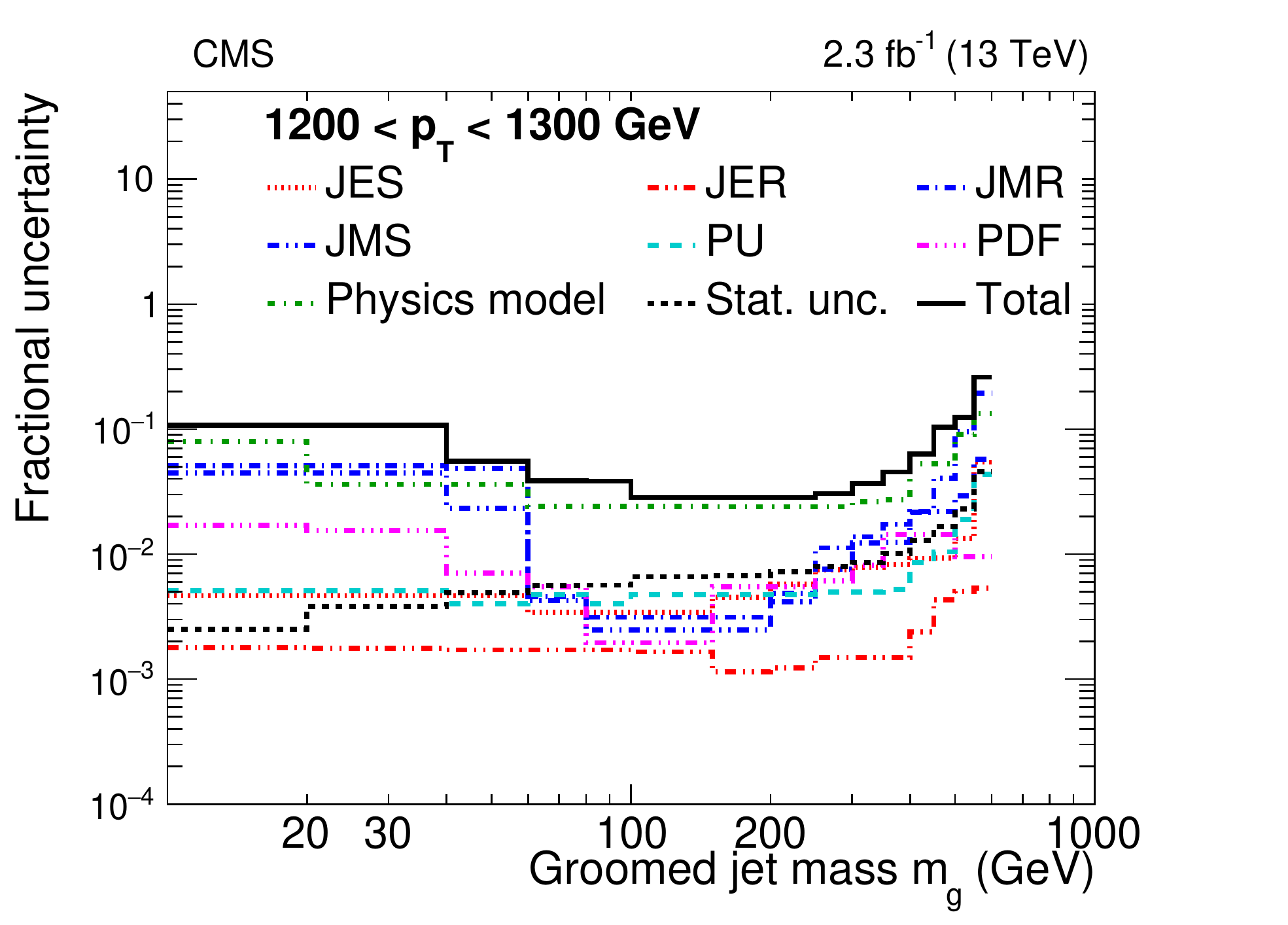}
\caption{Systematic uncertainties in the normalized cross section after unfolding for two $\pt$ bins
  before grooming (top) and after grooming (bottom). The red dotted lines show the JES uncertainties. The red dash-dot-dotted lines show the JER uncertainties. The blue dash-dotted lines show the JMR uncertainties. The blue dash-dot-dotted lines show the JMS uncertainties. The cyan dashed lines show the PU uncertainties. The orange dash-dotted lines show the luminosity uncertainties. The magenta dash-dot-dot-dotted lines show the PDF uncertainties. The green dash-dotted lines show the physics model uncertainties. The black short-dashed lines show the statistical uncertainties.} 
\label{fig:Unc1}
\end{figure}

\clearpage
\section{Results}

The absolute cross sections as a function of the jet mass are shown for two $\pt$ bins in
Fig.~\ref{fig:AbsResults1}.
The cross sections are compared to {\PYTHIA}8,
\POWHEG+ {\PYTHIA}8, and
\HERWIGpp predictions, which are scaled to match the number of events in the data.

The
normalized cross sections as a function of the jet mass are shown for
two $\pt$ bins in
Fig.~\ref{fig:Results1}.  
All $\pt$ bins of the normalized cross section are shown in Figs.~\ref{fig:fullstack}
and~\ref{fig:fullstacksoftdrop}.
The lowest masses in the measurement
are difficult to measure experimentally, as described in Section~\ref{sec:uncertainties}.
As mentioned there, the mass measurements are presented for ungroomed jets for $m>20$\GeV when
$\pt < 760$\GeV, and $m > 40$\GeV for jets with
$\pt > 760$\GeV. Results for groomed jets are presented for $m > 10$\GeV
for all $\pt$.
The normalized
results
for the groomed jets are also compared to the calculation
of the jet mass from Refs.~\cite{Frye:2016aiz}
and~\cite{Marzani:2017mva}. The normalized results are scaled
to unity for each individual jet $\pt$ bin for the unfolded data, for the
{\PYTHIA}8, \HERWIGpp, \POWHEG+ {\PYTHIA}8 predictions, and for the results from the
calculation from the authors of Ref.~\cite{Marzani:2017mva}. The
calculation from Ref.~\cite{Frye:2016aiz},
on the other hand, is normalized to match the
data at 50\GeV, since the calculation does not consider
nonperturbative corrections that are important below that value.
As a reminder, the prediction from Ref.~\cite{Frye:2016aiz} includes nonperturbative effects
as an uncertainty, while the prediction from Ref.~\cite{Marzani:2017mva}
directly includes nonperturbative effects in the prediction, and has smaller systematic
uncertainties in low-mass regions where nonperturbative effects are important.

Comparing the normalized ungroomed and groomed cross sections (Fig.~\ref{fig:Results1}), 
our results show that the grooming algorithm considerably lowers the jet mass and suppresses the Sudakov peak, as expected.   
The precision of the measurement improves, 
since the grooming algorithm removes the 
parts of the jet arising from soft radiation and the pileup contribution. 
Figure~\ref{fig:Unc1} shows the reduced effect of the uncertainties from the physics model and pileup,
decreasing the overall uncertainty from $\sim$8--12\% at $m=100$\GeV for ungroomed jets
to $\sim$2--5\% for groomed jets at the same mass.
The absolute cross sections (Fig.~\ref{fig:AbsResults1})
have larger uncertainties than the normalized cross sections because of jet energy uncertainties (JES and JER),
as shown in Fig.~\ref{fig:AbsUnc1}.

The LO theory predictions with a MC-based physics model
are from {\PYTHIA}8 and \HERWIGpp, and they
predict the ungroomed jet mass measurement within uncertainties, although the experimental uncertainties are large. Above
the splitting threshold ($m/\pt > 0.3$), the prediction is slightly larger than the data.
After grooming, the experimental precision improves, so some deviations from the data can be observed at
very low jet masses.
Some disagreements are observed
between {\PYTHIA}8 and \HERWIGpp. The {\PYTHIA}8
generator tends to predict the data better,
although for $m/\pt < 0.1$, the agreement is worse than in the range $0.1 < m/\pt < 0.3$.
The same disagreement above the splitting threshold is observed as in the ungroomed case.
The NLO theory prediction with a MC-based physics model is from
\POWHEG+ {\PYTHIA}8, and
is observed to have largely the same behavior as {\PYTHIA}8 alone, so the
largest effects are coming from the physics model.

The LO and NLO theory predictions with analytic resummation
(Refs.~\cite{Frye:2016aiz} and~\cite{Marzani:2017mva}, respectively) agree overall
with the data, with some slight disagreements at very low masses.
For $m/\pt > 0.3$, the fixed-order
matrix element merging is insufficient to capture the true
dynamics due to kinematic effects, as described in Section~\ref{sec:theory}, so the predictions diverge from the data.
They agree with the predictions using MC-based physics models, although
some disagreement is observed at very low masses.
The resummed predictions are slightly closer to the {\PYTHIA}8 prediction than to the \HERWIGpp one.

\begin{figure}[htb!]
\centering
\includegraphics[width=0.49\textwidth]{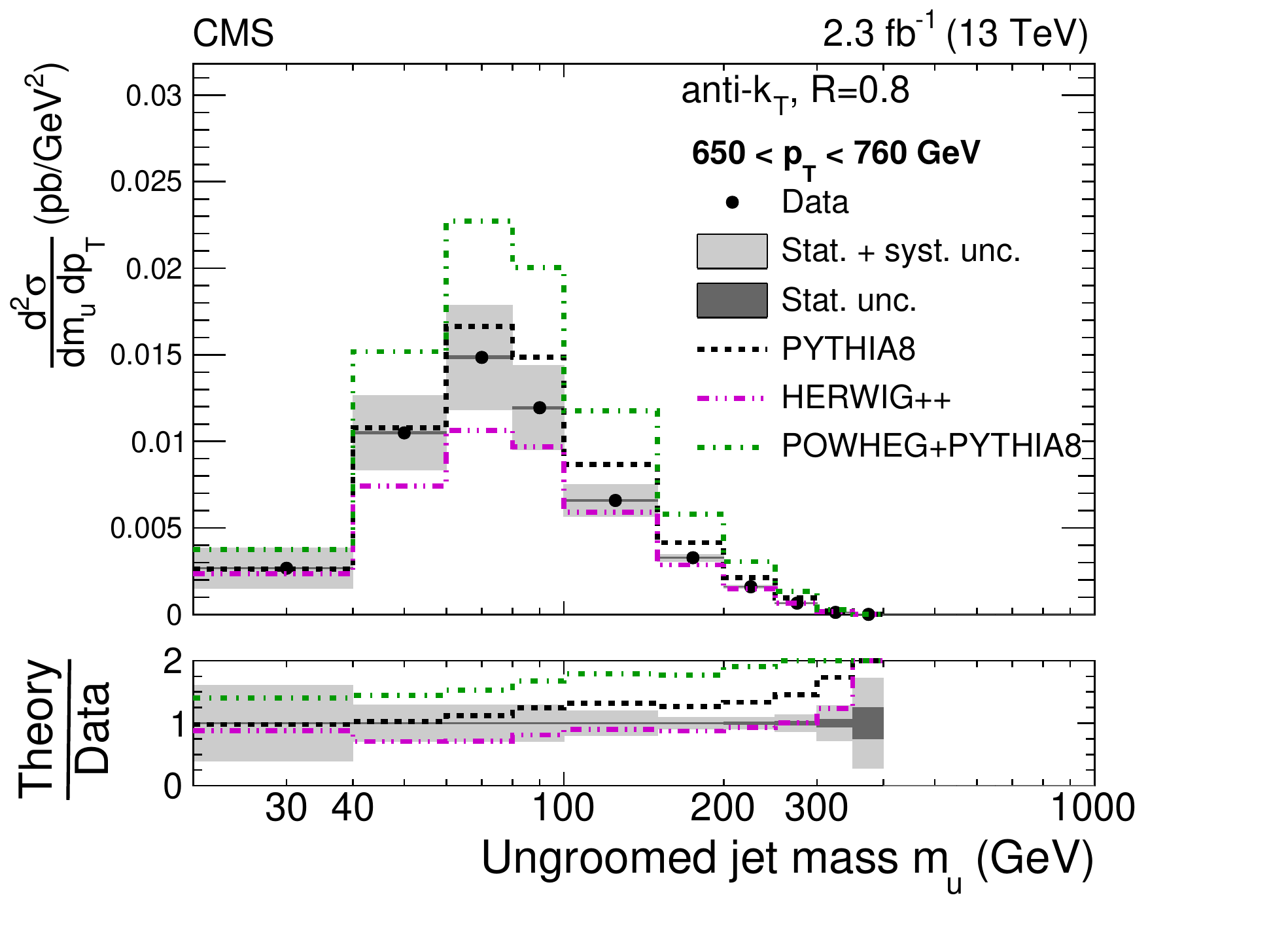}
\includegraphics[width=0.49\textwidth]{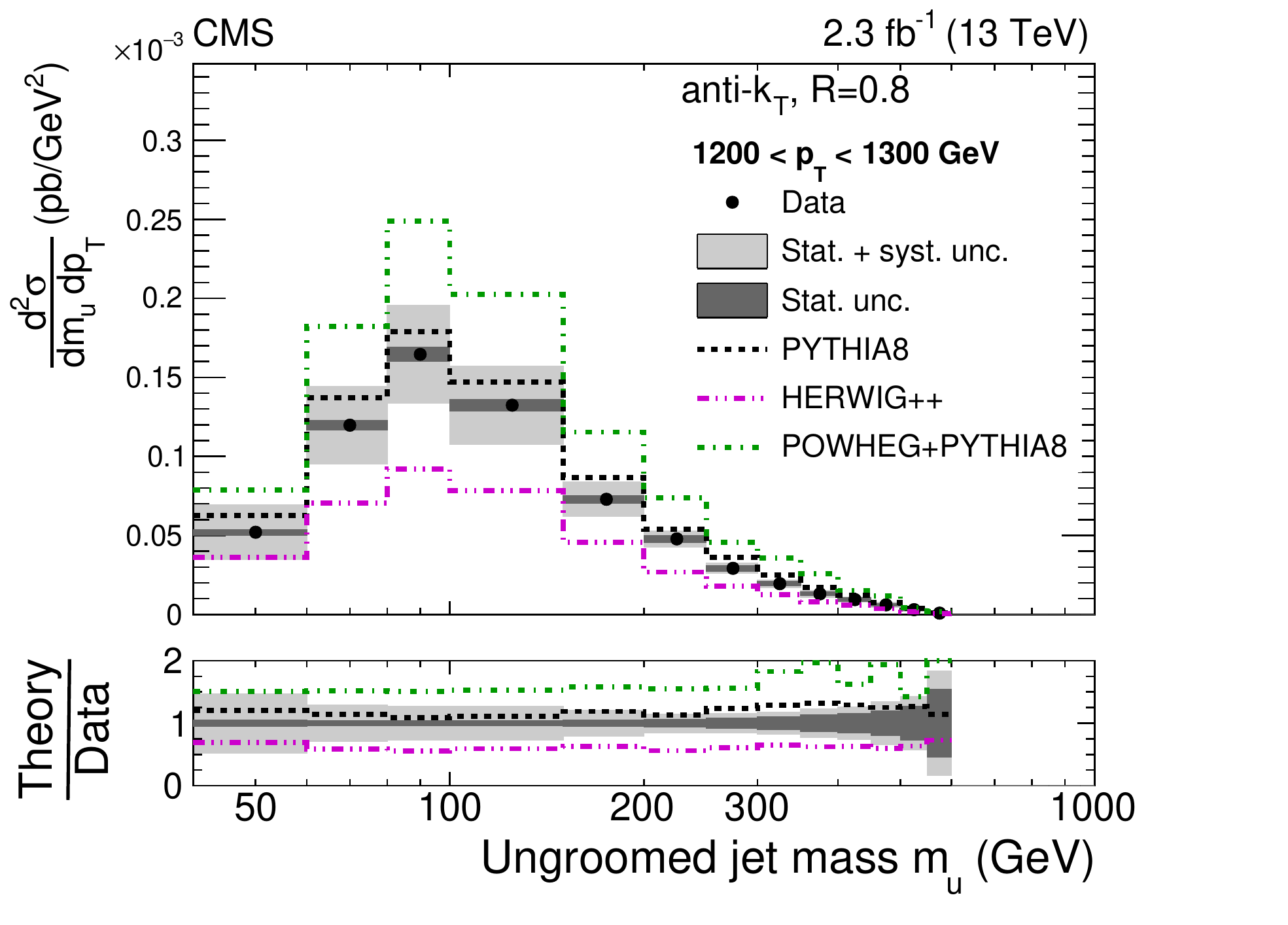}\\
\includegraphics[width=0.49\textwidth]{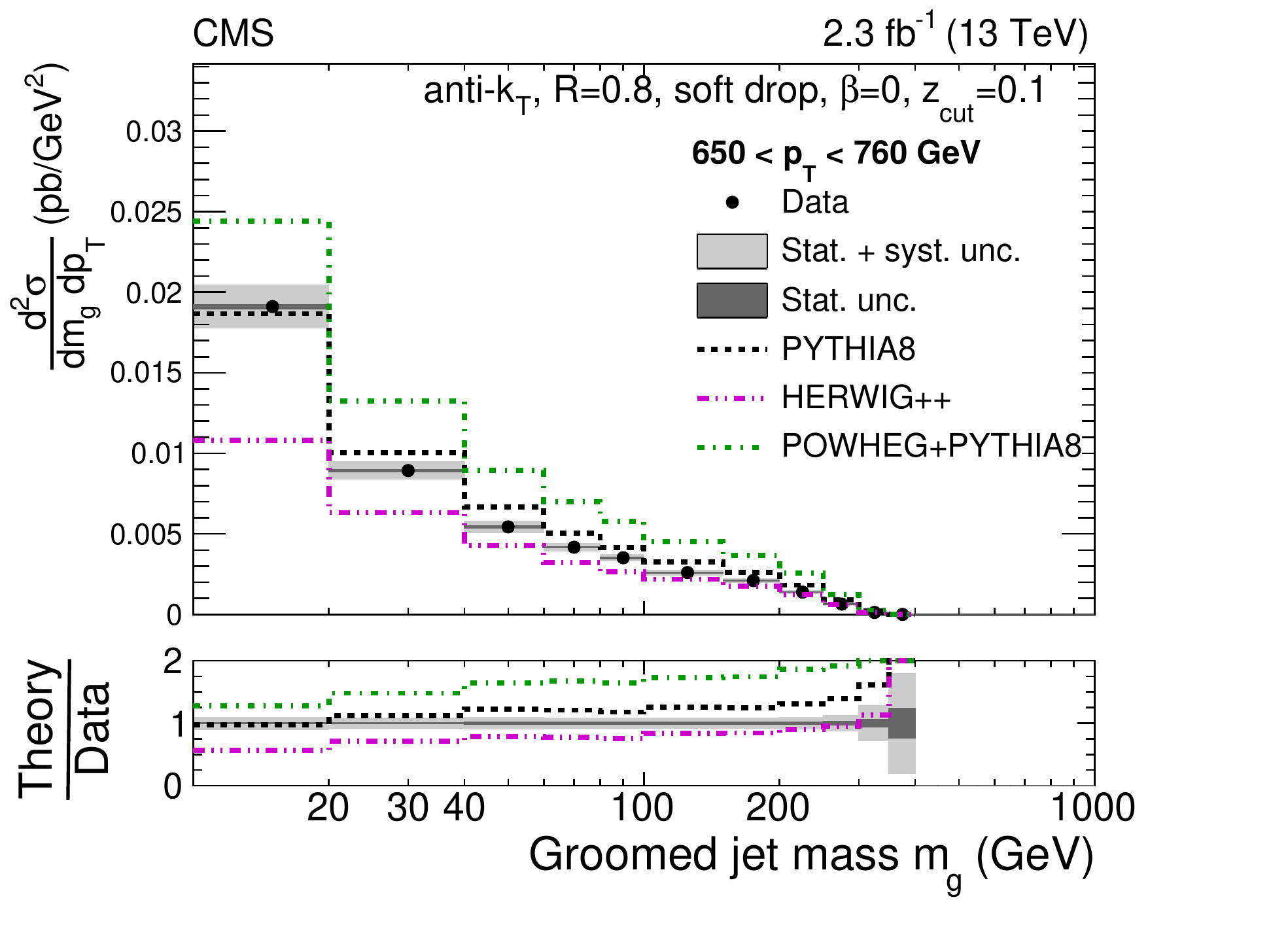}
\includegraphics[width=0.49\textwidth]{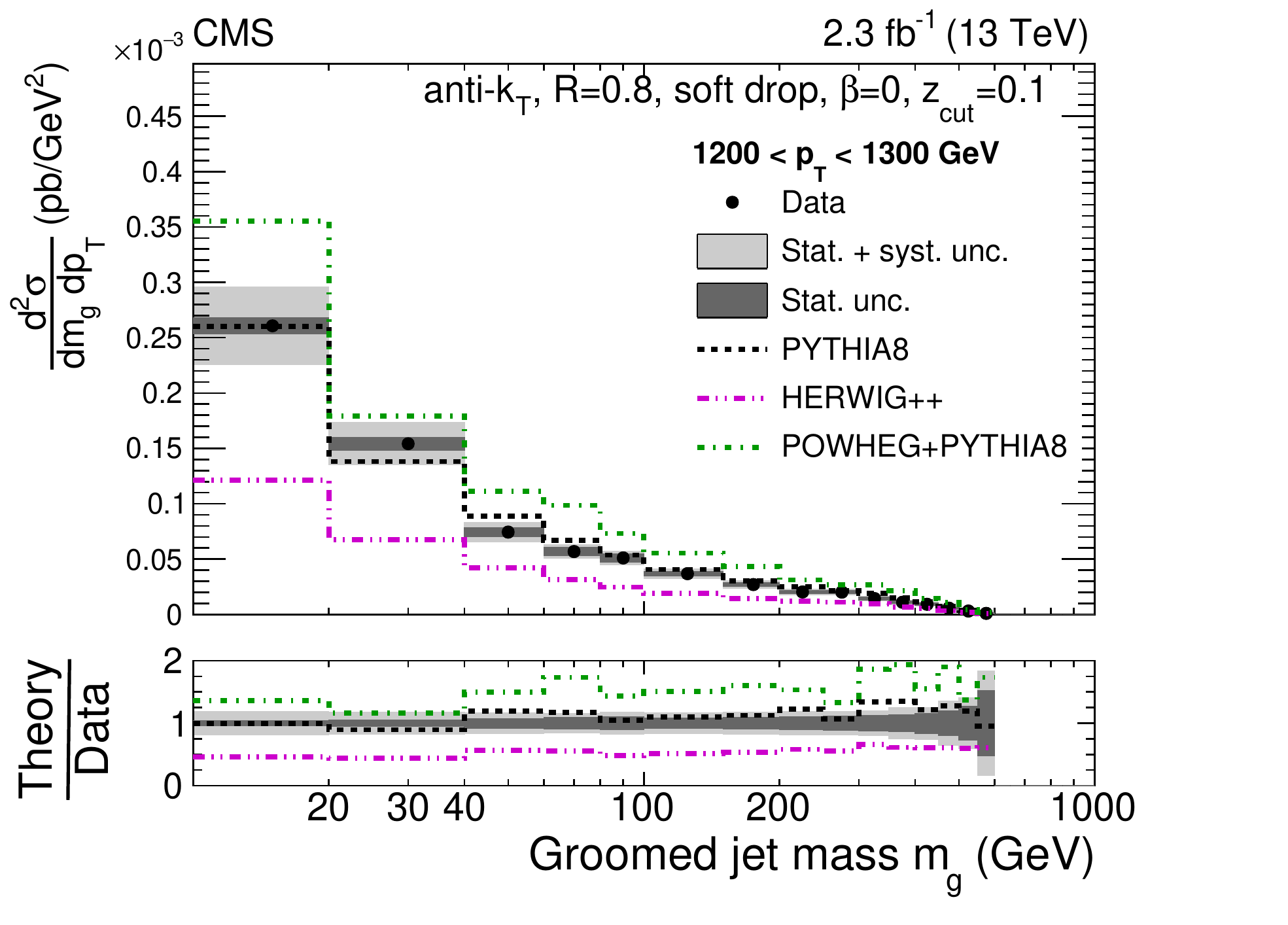}
\caption{Absolute cross section for ungroomed (top) and groomed (bottom) jets for two $\pt$ bins. The
  data are shown by the black points, with dark grey bands for the statistical uncertainty (Stat. unc.)
  and with light grey bands for the total uncertainty (Stat. +
  sys. unc., added in quadrature). The predictions
  from {\PYTHIA}8, \HERWIGpp, and \POWHEG+ \PYTHIA are shown with
  dashed black, dash-dot-dotted magenta, and dash-dotted green histograms, respectively,
  with no uncertainties shown. The
  MC simulations are normalized to the data over the entire $\pt$ range.}
\label{fig:AbsResults1}
\end{figure}

\begin{figure}[htb!]
\centering
\includegraphics[width=0.49\textwidth]{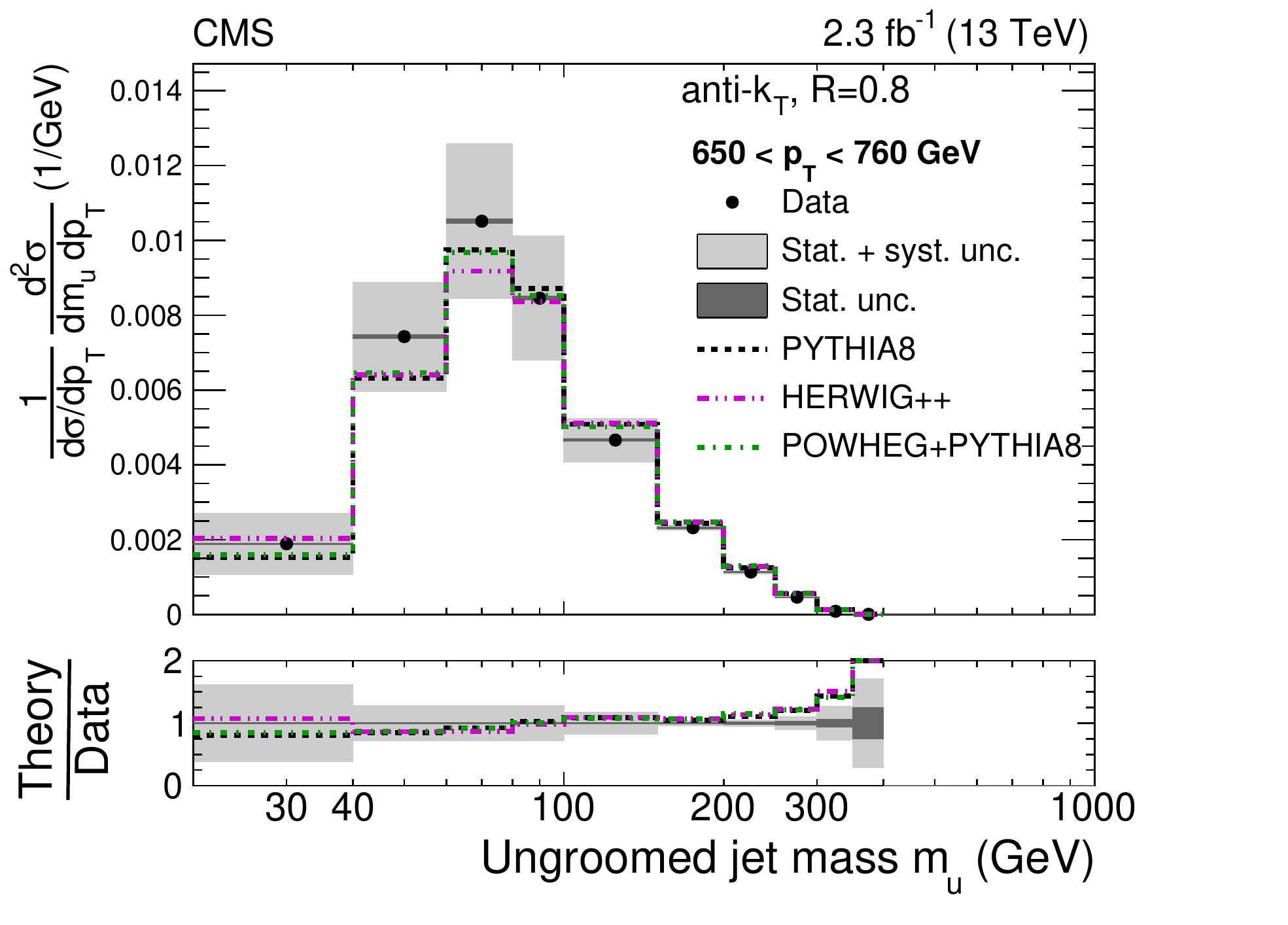}
\includegraphics[width=0.49\textwidth]{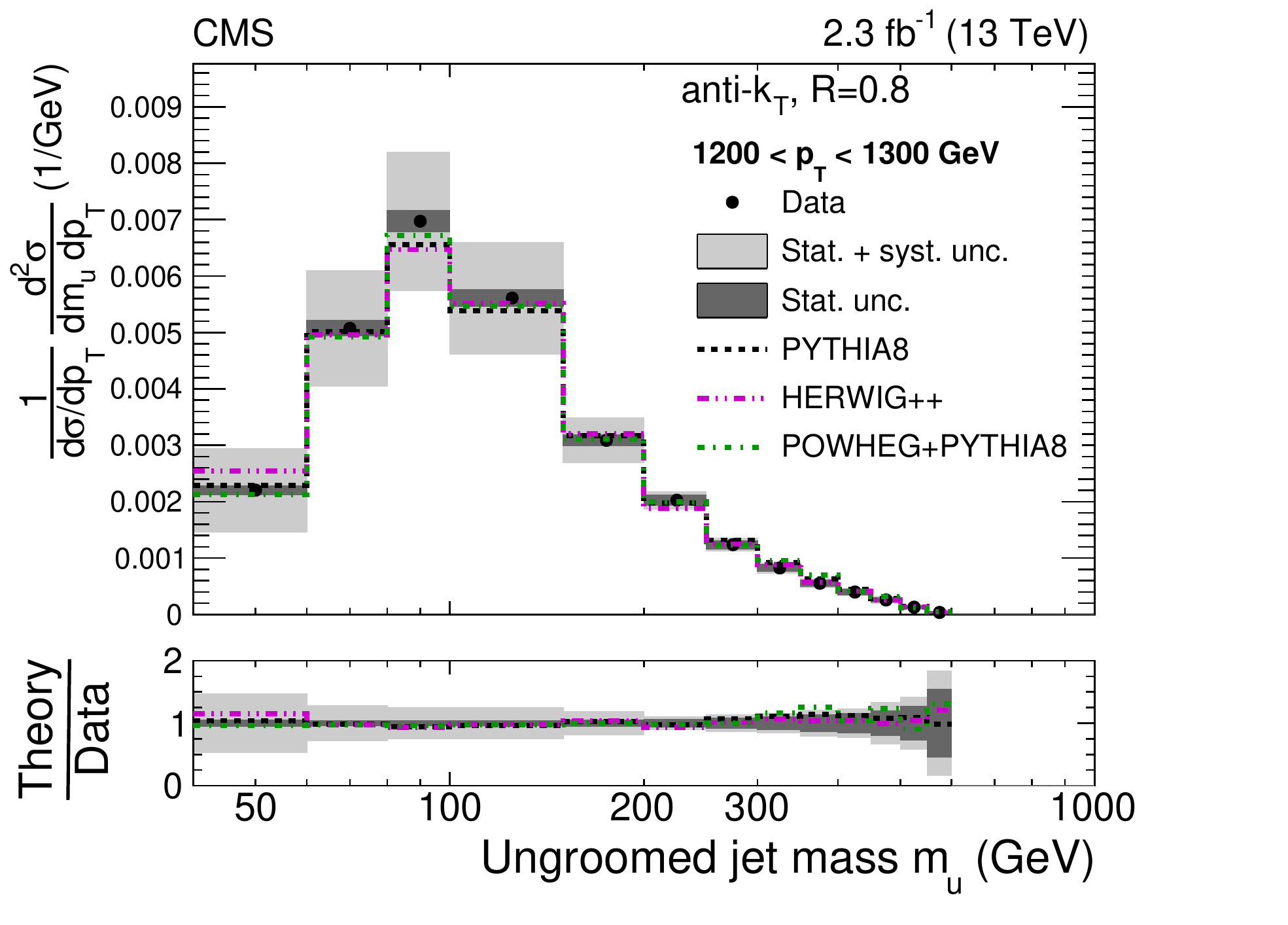}\\
\includegraphics[width=0.49\textwidth]{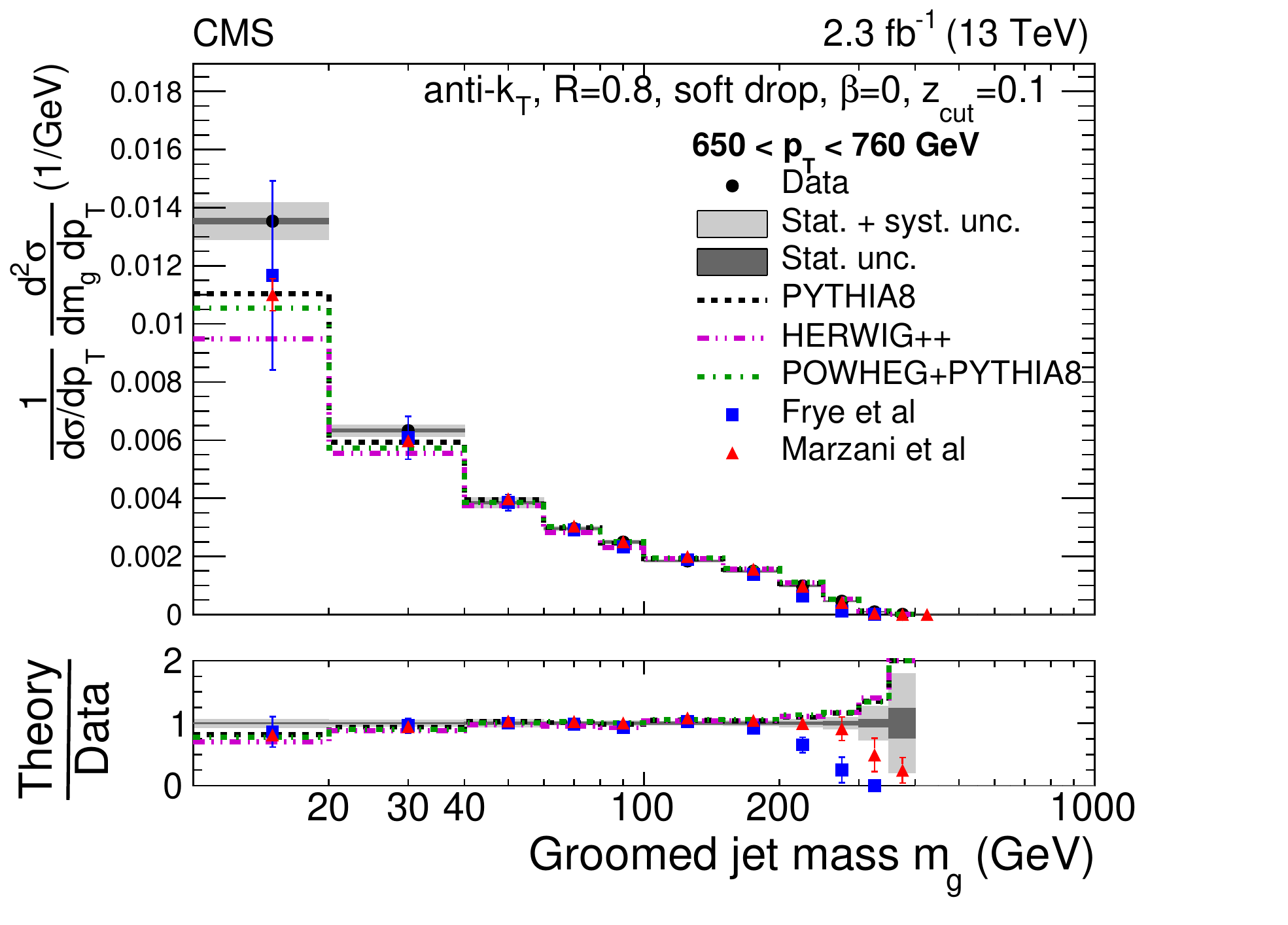}
\includegraphics[width=0.49\textwidth]{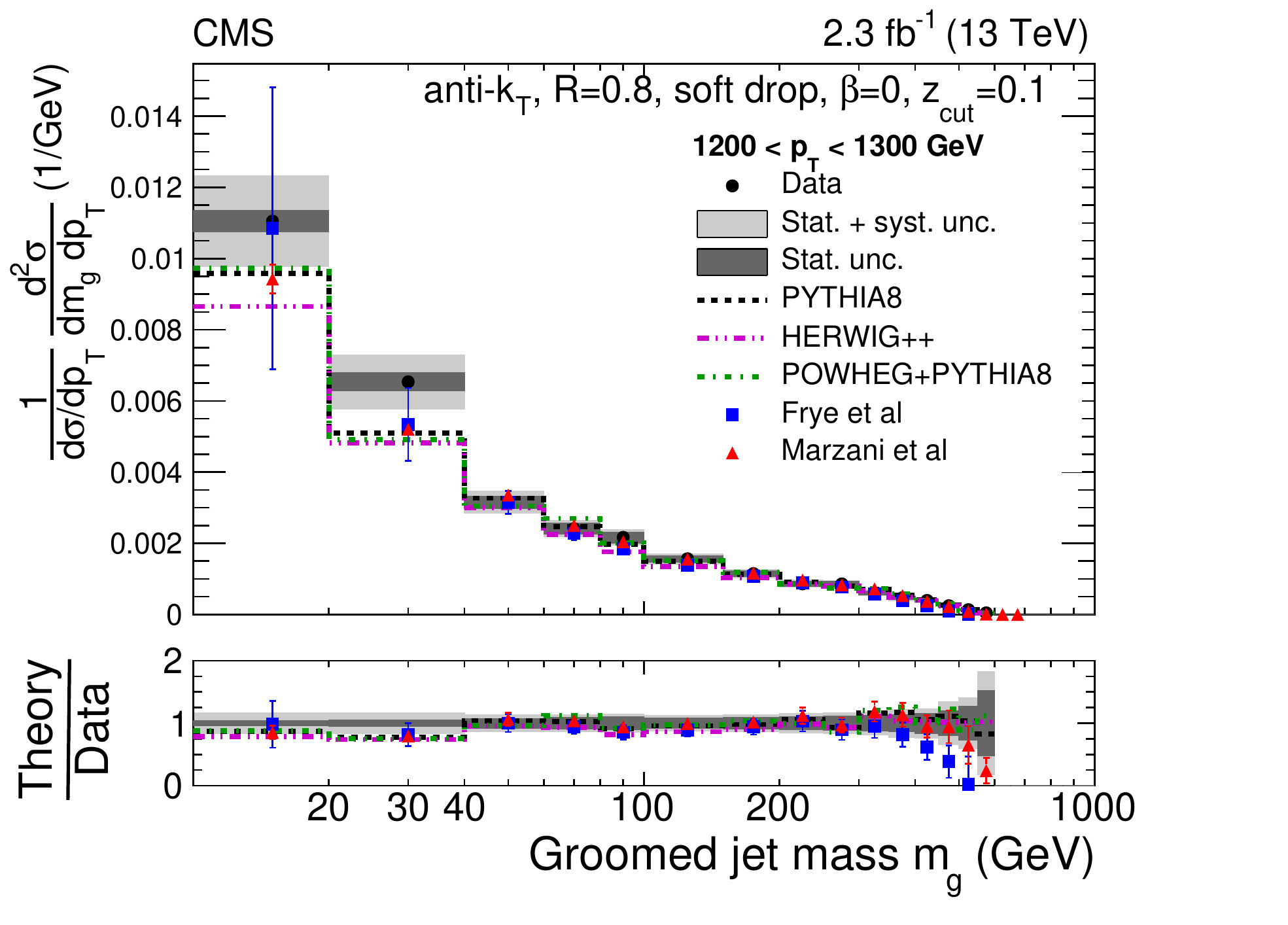}
\caption{Normalized cross section for the ungroomed (top) and groomed
  (bottom) jets for two $\pt$ bins. The predictions
  from {\PYTHIA}8, \HERWIGpp, and \POWHEG+ \PYTHIA are shown with
  dashed black, dash-dot-dotted magenta, and dash-dotted green histograms, respectively,
  with no uncertainties shown. 
  The predictions from Ref.~\cite{Frye:2016aiz} (Frye et al.) are shown with
  blue squares.
  The uncertainties include scale variations and an estimate of
  nonperturbative effects. The predictions from
  Ref.~\cite{Marzani:2017mva} (Marzani et al.) are shown with red triangles. 
  The uncertainties include only
  effects from scale variations, since nonperturbative corrections
  have been considered directly in the calculation. Both predictions
  diverge from the data at high mass due to fixed-order
  matching.}
\label{fig:Results1}
\end{figure}

\clearpage

\begin{figure}[htb!]
\centering
\includegraphics[width=0.8\textwidth]{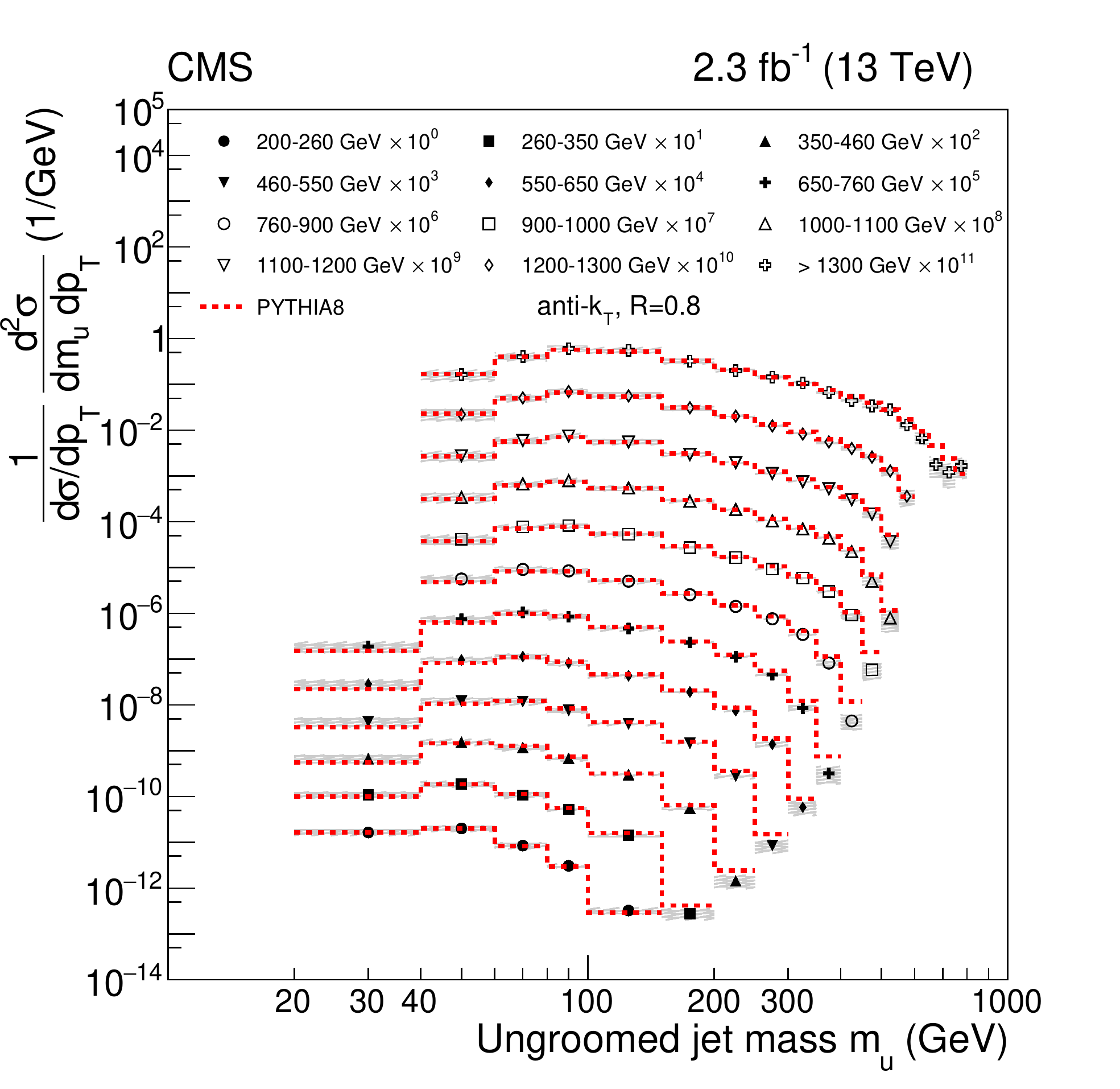}
\caption{Normalized cross section for ungroomed jets for all $\pt$ bins. Bins
  with statistical uncertainty larger than 60\% are not shown. The
  data are shown with markers for each $\pt$ bin, scaled by a factor for
  better visibility. The total uncertainties
  (statistical added to systematic in quadrature) are shown with grey bands. The
  predictions from {\PYTHIA}8 are shown as a dashed red line.}
\label{fig:fullstack}
\end{figure}

\begin{figure}[htb!]
\centering
\includegraphics[width=0.8\textwidth]{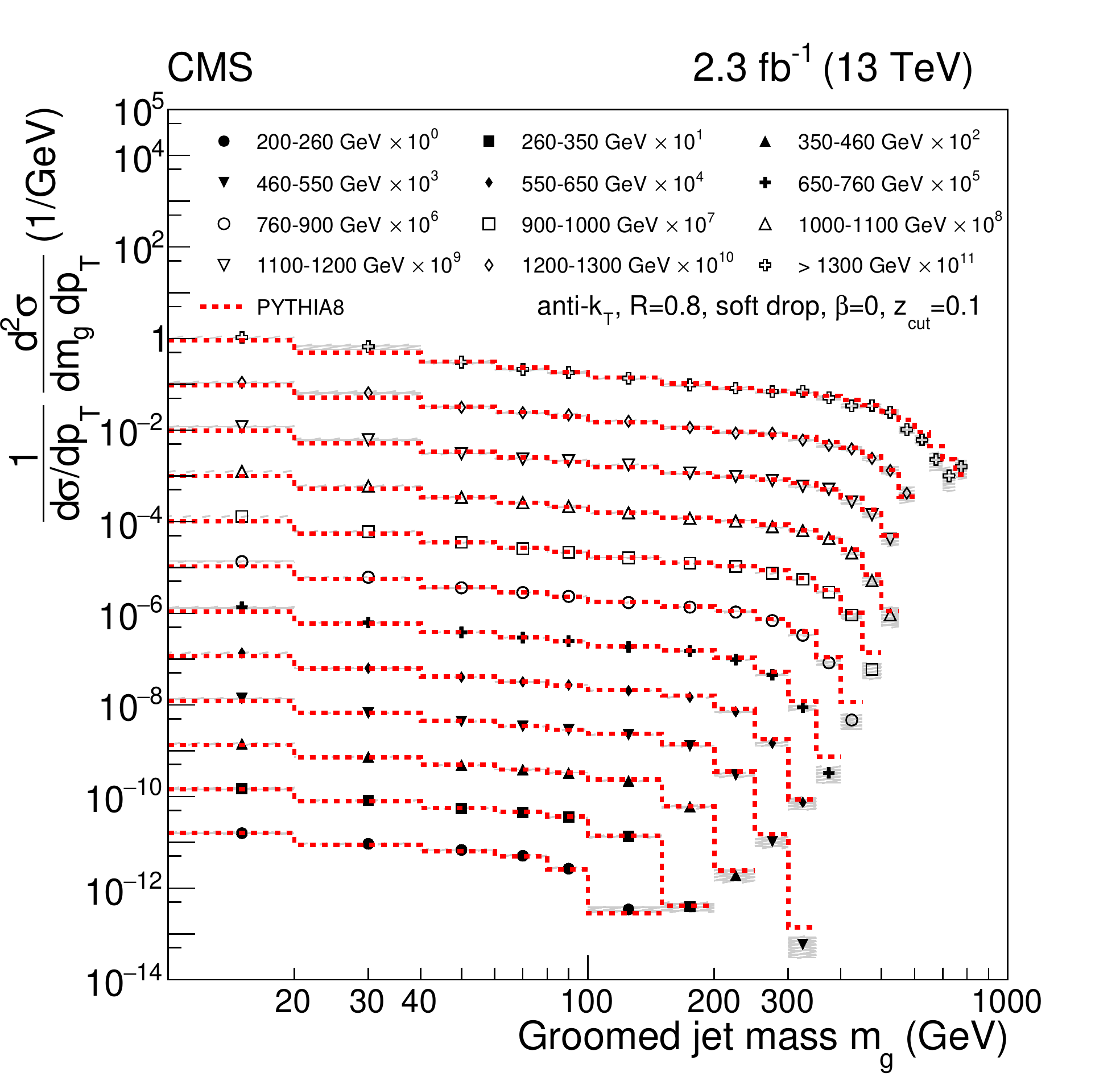}
\caption{Normalized cross section for groomed jets for all $\pt$ bins. Bins
  with statistical uncertainty larger than 60\% are not shown. The
  data are shown with markers for each $\pt$ bin, scaled by a factor for
  better visibility. The total uncertainties
  (statistical added to systematic in quadrature) are shown with grey bands. The
  predictions from {\PYTHIA}8 are shown as a dashed red line. }
\label{fig:fullstacksoftdrop}
\end{figure}

\section{Summary}

The double-differential production cross section has been presented for
dijet events in bins of ungroomed jet $\pt$
as a function of the jet mass, with and without a jet grooming algorithm applied. 
The grooming algorithm is observed to decrease the jet mass overall, and reduce
the sensitivity of the observable to details of the physics modeling
and pileup effects.
The data are compared to various theoretical predictions, including new calculations
using analytical resummation.
Overall, in the normalized cross section, the theoretical predictions
agree with the measured cross sections within the uncertainties for
masses from 10 to 30\% of the jet transverse momentum.

\clearpage

\begin{acknowledgments}
We congratulate our colleagues in the CERN accelerator departments for the excellent performance of the LHC and thank the technical and administrative staffs at CERN and at other CMS institutes for their contributions to the success of the CMS effort. In addition, we gratefully acknowledge the computing centres and personnel of the Worldwide LHC Computing Grid for delivering so effectively the computing infrastructure essential to our analyses. Finally, we acknowledge the enduring support for the construction and operation of the LHC and the CMS detector provided by the following funding agencies: BMWFW and FWF (Austria); FNRS and FWO (Belgium); CNPq, CAPES, FAPERJ, FAPERGS, and FAPESP (Brazil); MES (Bulgaria); CERN; CAS, MoST, and NSFC (China); COLCIENCIAS (Colombia); MSES and CSF (Croatia); RPF (Cyprus); SENESCYT (Ecuador); MoER, ERC IUT, and ERDF (Estonia); Academy of Finland, MEC, and HIP (Finland); CEA and CNRS/IN2P3 (France); BMBF, DFG, and HGF (Germany); GSRT (Greece); NKFIA (Hungary); DAE and DST (India); IPM (Iran); SFI (Ireland); INFN (Italy); MSIP and NRF (Republic of Korea); MES (Latvia); LAS (Lithuania); MOE and UM (Malaysia); BUAP, CINVESTAV, CONACYT, LNS, SEP, and UASLP-FAI (Mexico); MOS (Montenegro); MBIE (New Zealand); PAEC (Pakistan); MSHE and NSC (Poland); FCT (Portugal); JINR (Dubna); MON, RosAtom, RAS, RFBR, and NRC KI (Russia); MESTD (Serbia); SEIDI, CPAN, PCTI, and FEDER (Spain); MOSTR (Sri Lanka); Swiss Funding Agencies (Switzerland); MST (Taipei); ThEPCenter, IPST, STAR, and NSTDA (Thailand); TUBITAK and TAEK (Turkey); NASU and SFFR (Ukraine); STFC (United Kingdom); DOE and NSF (USA).

\hyphenation{Rachada-pisek} Individuals have received support from the Marie-Curie programme and the European Research Council and Horizon 2020 Grant, contract No. 675440 (European Union); the Leventis Foundation; the A. P. Sloan Foundation; the Alexander von Humboldt Foundation; the Belgian Federal Science Policy Office; the Fonds pour la Formation \`a la Recherche dans l'Industrie et dans l'Agriculture (FRIA-Belgium); the Agentschap voor Innovatie door Wetenschap en Technologie (IWT-Belgium); the F.R.S.-FNRS and FWO (Belgium) under the ``Excellence of Science - EOS" - be.h project n. 30820817; the Ministry of Education, Youth and Sports (MEYS) of the Czech Republic; the Lend\"ulet (``Momentum") Programme and the J\'anos Bolyai Research Scholarship of the Hungarian Academy of Sciences, the New National Excellence Program \'UNKP, the NKFIA research grants 123842, 123959, 124845, 124850 and 125105 (Hungary); the Council of Science and Industrial Research, India; the HOMING PLUS programme of the Foundation for Polish Science, cofinanced from European Union, Regional Development Fund, the Mobility Plus programme of the Ministry of Science and Higher Education, the National Science Center (Poland), contracts Harmonia 2014/14/M/ST2/00428, Opus 2014/13/B/ST2/02543, 2014/15/B/ST2/03998, and 2015/19/B/ST2/02861, Sonata-bis 2012/07/E/ST2/01406; the National Priorities Research Program by Qatar National Research Fund; the Programa Estatal de Fomento de la Investigaci{\'o}n Cient{\'i}fica y T{\'e}cnica de Excelencia Mar\'{\i}a de Maeztu, grant MDM-2015-0509 and the Programa Severo Ochoa del Principado de Asturias; the Thalis and Aristeia programmes cofinanced by EU-ESF and the Greek NSRF; the Rachadapisek Sompot Fund for Postdoctoral Fellowship, Chulalongkorn University and the Chulalongkorn Academic into Its 2nd Century Project Advancement Project (Thailand); the Welch Foundation, contract C-1845; and the Weston Havens Foundation (USA).

We would like to thank Christopher Frye, Andrew
Larkoski, Simone Marzani, Gavin Salam, Matthew Schwartz, Lais Schunk,
Gregory Soyez, and Kai Yan for providing calculations of theoretical predictions of the jet mass distribution for this analysis.
\end{acknowledgments}

\bibliography{auto_generated}

\cleardoublepage \appendix\section{The CMS Collaboration \label{app:collab}}\begin{sloppypar}\hyphenpenalty=5000\widowpenalty=500\clubpenalty=5000\vskip\cmsinstskip
\textbf{Yerevan Physics Institute, Yerevan, Armenia}\\*[0pt]
A.M.~Sirunyan, A.~Tumasyan
\vskip\cmsinstskip
\textbf{Institut f\"{u}r Hochenergiephysik, Wien, Austria}\\*[0pt]
W.~Adam, F.~Ambrogi, E.~Asilar, T.~Bergauer, J.~Brandstetter, M.~Dragicevic, J.~Er\"{o}, A.~Escalante~Del~Valle, M.~Flechl, R.~Fr\"{u}hwirth\cmsAuthorMark{1}, V.M.~Ghete, J.~Hrubec, M.~Jeitler\cmsAuthorMark{1}, N.~Krammer, I.~Kr\"{a}tschmer, D.~Liko, T.~Madlener, I.~Mikulec, N.~Rad, H.~Rohringer, J.~Schieck\cmsAuthorMark{1}, R.~Sch\"{o}fbeck, M.~Spanring, D.~Spitzbart, A.~Taurok, W.~Waltenberger, J.~Wittmann, C.-E.~Wulz\cmsAuthorMark{1}, M.~Zarucki
% ...
\vskip\cmsinstskip
% ...
1:  Also at Vienna University of Technology, Vienna, Austria\\
% ...
\end{sloppypar}
\end{document}